\documentclass{aa}  

\usepackage{graphicx}
\usepackage{txfonts}
\def\kms{\,{\rm km}\,{\rm s}^{-1}}

\def\feh{\hbox{[Fe/H]}}
\def\mgfe{\hbox{[Mg/Fe]}}
\def\tife{\hbox{[Ti/Fe]}}
\def\mnfe{\hbox{[Mn/Fe]}}
\def\teff{T_{\rm eff}}
\def\Vmic{V_{\rm mic}}

\def\Vbrd{V_{\rm broad}}
\def\logg{\log({\rm g})}
\def\snr{\hbox{S/N}}
\begin{document}

\title{NLTE Chemical abundances in Galactic open and globular clusters\thanks{Based  on  data products from observations made with ESO Telescopes at the La Silla Paranal Observatory under programme ID 188.B-3002}.}
\subtitle{}
\titlerunning{Fe, Mg, Ti in the galactic clusters.}
\author{Mikhail Kovalev\inst{1}
          \and
          Maria Bergemann\inst{1}
          \and
          Yuan-Sen Ting\inst{2,3,4}\fnmsep\thanks{Hubble Fellow}
          \and
          Hans-Walter Rix\inst{1}
          }
\authorrunning{M. Kovalev et al.}
\institute{Max Planck Institute for Astronomy, K\"{o}nigstuhl 17, D-69117 Heidelberg, Germany\\
              \email{kovalev@mpia.de}
        \and
            Institute for Advanced Study, Princeton, NJ 08540, USA
        \and
            Department of Astrophysical Sciences, Princeton University, Princeton, NJ 08544, USA
        \and
            Observatories of the Carnegie Institution of Washington, 813 Santa Barbara Street, Pasadena, CA 91101, USA
             }

   \date{Received 9 May, 2019; accepted xxx}

 
\abstract
{}
{We study the effects of non-local thermodynamic equilibrium (NLTE) on the determination of stellar parameters and abundances of Fe, Mg, and Ti from the medium-resolution spectra of FGK stars.}
{We extend \textit{the Payne} fitting approach to draw on NLTE and LTE spectral models. These are used to analyse the spectra of the Gaia-ESO benchmark stars and the spectra of 742 stars in 13 open and globular clusters in the Milky Way: NGC 3532, NGC 5927, NGC 2243, NGC 104, NGC 1851, NGC 2808, NGC 362, M2, MGC 6752, NGC 1904, NGC 4833, NGC 4372 and M15.}
{Our approach accurately recovers effective temperatures, surface gravities, and abundances of the benchmark stars and clusters members. The differences between NLTE and LTE are significant in the metal-poor regime, [Fe/H] $\lesssim -1$. The NLTE $\feh$ values are systematically higher, whereas the average NLTE [Mg/Fe] abundance ratios are $\sim 0.15$ dex lower, compared to LTE. Our LTE measurements of metallicities and abundances of stars in Galactic clusters are in a good agreement with the literature. Yet, for most clusters, our study yields the first estimates of NLTE abundances of Fe, Mg and Ti.}
%
{All clusters investigated in this work are homogeneous in Fe and Ti, with the intra-cluster abundance variations of less then 0.04 dex. NGC 2808, NGC 4833, M2 and M 15 show significant dispersions in [Mg/Fe]. Contrary to common assumptions, the NLTE analysis changes the mean abundance ratios in the clusters, but it does not influence the intra-cluster abundance dispersions.}

\maketitle
%

\section{Introduction}

Fast and reliable modelling of stellar spectra is becoming increasingly important for current stellar and Galactic astrophysics. Large-scale spectroscopic stellar surveys, such as Gaia-ESO \citep{Gilmore2012,Randich2013}, APOGEE \citep{Majewski2015}, and GALAH \citep{DeSilva2015} are revolutionising our understanding of the structure and evolution of the Milky Way galaxy, stellar populations, and stellar physics. The ever-increasing amount of high-quality spectra, in return, demands rigorous, physically-realistic, and efficient data analysis techniques to provide an accurate diagnostic of stellar parameters and abundances. This problem has two sides. Precise spectral fitting and analysis requires powerful numerical optimisation and data-model comparison algorithms. On the other hand, the accuracy of stellar label estimates is mostly limited by the physics of spectral models used in the model-data comparison. The fitting aspect has been the subject of extensive studies over the past years, and various methods \citep[e.g.][]{matisse,Schoenrich2014,ness2015,CaseyCannonRAVE,ting2018} have been developed and applied to the analysis of large survey datasets.
\par
Major developments have also occurred in the field of stellar atmosphere physics. Non-local Thermodynamic Equilibrium (NLTE) radiative transfer is now routinely performed for many elements in the periodic table. This allows detailed calculations of spectral profiles that account for NLTE effects. NLTE models consistently describe the interaction of the gas particles in stellar atmospheres with the radiation field \citep{Auer1969a}, in this respect being more realistic than LTE models. In NLTE, photons affect atomic energy level populations, whilst in LTE those are set solely by the Saha equation for ionisation  and by the Boltzmann distribution for excitation. NLTE models predict more realistic absorption line profiles and hence provide more accurate stellar parameters and abundances \citep[e.g.][]{ruchti2013,zhao2016}. However NLTE models are often incomplete in terms of atomic data, such as collisions with H atoms and electrons or  photo-ionisation cross-sections. Major efforts to improve atomic data are underway \citep[e.g.][]{yakovleva2016, bautista2017, belyaev2017, barklem2017, amarsi2018, barklem2018} and there is no doubt that many gaps in the existing atomic and molecular databases will be filled in the near-term future. Besides, strictly speaking, no single NLTE model is complete in terms of atomic data, and also quantum-mechanical cross-sections are usually available for a small part of the full atomic or molecular system \citep{barklem2016}.

In this work, we study the effect of NLTE on the analysis of stellar parameters and chemical abundances for FGK-type stars. We combine NLTE stellar spectral models with \textit{the Payne}\footnote{\url{https://github.com/tingyuansen/The_Payne}} code developed by \citet{ting2018} and apply our methods to the observed stellar spectra from the 3$^{rd}$ public data release by the Gaia-ESO survey. This work is a proof-of-concept of the combined NLTE-Payne approach and it is, hence, limited to the analysis to the Gaia-ESO benchmark stars and a sample of Galactic open and globular clusters, for which independent estimates of stellar labels, both stellar parameters and detailed abundances are available from earlier studies.

The paper is organised as follows. In Section \ref{Method}, we describe the observed sample, the physical aspects of the theoretical spectral models, and the mathematical basis of  \textit{the Payne} code. We present the LTE and NLTE results in Section \ref{Results} and compare them with the literature in Section \ref{discussion}. Section \ref{Conclusions} summarises the conclusions and outlines future prospects arising from this work. 
%
%
\section{Methods}
\label{Method}

\subsection{Observed spectra}
\label{Observations}

We use the spectra of FGK stars observed within the Gaia-ESO spectroscopic survey \citep{Gilmore2012,Randich2013}. These spectra are now publicly available as a part of the third data release (DR3.1)\footnote{\url{http://archive.eso.org/wdb/wdb/adp/phase3_spectral/form?collection_name=GAIAESO}}. The data were obtained with the Giraffe instrument \citep{Pasquini2002} at the ESO (European Southern Observatory) VLT (Very Large Telescope). We use the spectra taken with the HR10 setting, which covers 280 \AA~ from 5334 \AA~ to 5611 \AA, at a resolving power of $R=\lambda/ \Delta \lambda \sim 19\,800$.  The average signal-to-noise ratio ($\snr$) of a spectrum ranges from 90 to 2800 per \AA\footnote{We employ the following relationship: $\snr$ [\AA$^{-1}$]=$\sqrt{20}~\snr$ [pixel$^{-1}$], where 20 pixels are equivalent to 1 \AA, that is, the sampling of the Giraffe HR10 spectra.}, with the majority of the spectra sampling the $\snr$ in range of 150-200 \AA$^{-1}$.

Our observed sample contains 916 FGK-type stars with luminosity classes from  III to V that includes main-sequence (MS), subgiants, and red giant branch (RGB) stars. A fraction of these are the Gaia-ESO benchmark stars (174 spectra of 19 stars), but we also include 742 stars in two open and 11 globular clusters. We exclude four benchmark stars with effective temperature $\teff<4000$~K, because this regime of stellar parameters is not covered by our model atmosphere grids. $\beta$~Ara is not a part of our calibration sample, as it is not recommended as a benchmark in \citet{Pancino2017}. These stars are previously analysed by Gaia-ESO \citep{Smiljanic2014a,sanroman2015,Pancino2017a} and included in the The Gaia-ESO DR3 catalogue. 

We estimate the radial velocity (RV) by cross-correlating the observed spectrum with a synthetic metal-poor  spectral template ($\teff=5800$~K,$~\logg=4.5$~dex, $ \feh=-2$~ dex)\footnote{Our tests showed that this template provides robust RV estimates for the full metallicity range.}, which is shifted in the RV range of $\pm 400~\kms$ with a step of $0.5~\kms$. We compute the cross-correlation function for all RV values and fit a parabola to $20$ points around the maximum value of the cross-correlation function. Then we apply the Doppler-shift to the observed spectrum using the velocity value at the position of the peak of the parabola. Since cross-correlation can incur small errors due to step size/template choice, we later fit for residual shift in the range $\pm~2 \kms$. %
\subsection{Model atmospheres and synthetic spectra}
The grids of LTE and NLTE synthetic spectra are computed using the new online spectrum synthesis tool \url{http://nlte.mpia.de}. The model atmospheres are 1D plane-parallel hydrostatic LTE models taken from the MAFAGS-OS grid \citep{Grupp2004,Grupp2004a}. For the NLTE grid we first compute the NLTE atomic number densities for Mg~\citep{Bergemann2017a}, Ti \citep{Bergemann2011}, Fe \citep{Bergemann2012c} and Mn \citep{bergemann2008} using the DETAIL statistical equilibrium (SE) code \citep{detail}. These are then fed into the SIU \citep{SIUReetz} radiative transfer (RT) and spectrum synthesis code. In total, $626$ spectral lines of Mg I, Ti I, Fe I and Mn I are modelled in NLTE for the NLTE grid, while for the LTE grid these lines are modelled with default LTE atomic level populations. Our approach is conceptually similar to \citet{buder2018a}, but we employ different SE and RT codes. We have chosen to use the MAFAGS-OS atmosphere grids, because these are internally consistent with DETAIL and SIU. In particular, the latter codes adopt the atomic and molecular partial pressures and partition functions that are supplied with the MAFAGS-OS models.

We compute $20\,000$ spectral models with $\teff$ uniformly distributed in the range from $4000$ to $7000$ K and $\logg$s in the range from $1.0$ to $5.0$ dex.  Metallicity\footnote{Hereafter, the abundance of iron $\feh$, is used as a proxy for metallicity.}, $\feh$, is uniformly distributed in the range from $\feh = -2.6$ to $0.5$ dex. We also allow for random variations in the rations of the magnesium, titanium, manganese to iron: [Mg/Fe], [Ti/Fe] from $-0.4$ to $0.8$ dex and [Mn/Fe] from $-0.8$ to $0.4$~dex. The abundances of other chemical elements are assumed to be solar and follow the iron abundance $\feh$.  In the metal-poor regime ($\feh<-1$ dex), some elements (like important opacity contributors C and O) can be significantly enhanced relative to the solar values. Therefore we computed several metal-poor synthetic spectra using a $0.5$ dex enhancement of C and O abundances and found that there is no impact on the spectral models. Micro-turbulence varies from 0.6 to 2.0 $\kms$, in line with high-resolution studies of FGK stars \citep[e.g.][]{ruchti2013}. The detailed solar abundances assumed in the MAFAGS-OS grids are reported in \citet{Grupp2004}. For the elements treated in NLTE, we adopt logA(Mg)$_\odot=7.58$ dex, logA(Ti)$_\odot=4.94$ dex, logA(Mn)$_\odot=5.53$ dex and logA(Fe)$_\odot=7.50$ dex \citep[meteoritic values from][]{grevesse1998}.

The widths of spectral lines in the observed spectra depend on many effects, such as the properties of the instrument, turbulence in stellar atmospheres, and stellar rotation \citep{gray}. However, it is not possible to separate these effects at the resolution of the Giraffe spectra. Hence, the macroturbulence, $V_{\rm mac}$, and the projected rotation velocity, $V{\rm sin~i}$, are dealt with by smoothing the model spectra with a Gaussian kernel, which corresponds to a characteristic velocity $V_{\rm broad}$ in the range from $5.0$ to $25.0$ $\kms$ that encompasses the typical values of $V_{\rm mac}$ and $V{\rm sin~i}$ reported for FGK stars \citep{gray,Jofre2015}. After that, the synthetic spectra are degraded to the resolution of the HR10 setup by convolving them with an instrumental profile (Appendix \ref{LSF})  and are re-sampled onto the observed spectrum wavelength grid using the sampling of $20$ wavelength points per \AA.
\subsection{The Payne code}
\label{payne}
The data-model comparison is not performed directly. Instead, we use \textit{the Payne} code to interpolate in the grid of synthetic spectra.

The approach consists of two stages: the training (model building) and the test (data fitting) steps. In the training step, we build a \textit{Payne} model using a set of pre-computed LTE and NLTE stellar spectra. We approximate the variation of the flux using an artificial neural network (ANN). In the test step, $\chi^2$ minimisation is employed to find the best-fit stellar parameters and abundances by comparing the model spectra to the observations. In what follows, we describe the key details of the method. For more details on the algorithm, we refer the reader to \citet{ting2018}.

The conceptual idea of the code is simple. We employ a simple ANN that consists of several fully connected layers of neurons: an input layer, two hidden layers, and an output layer. The input data are given by a set of stellar parameters (hereafter, labels) $\teff$, $\logg$, $\Vmic$, $V_{\rm broad}$, $\feh$, $\mgfe$, $\tife$ and $\mnfe$. The output data comprise the normalised flux values tabulated on a wavelength grid, as a function of the input labels.
Three hundred neurons in each hidden layer apply a weight and an offset to the output from the previous layer, and these outputs are activated using a $ReLU(z)={\rm max}(z,0)$ function for the first layer and a sigmoid function $s(z)=(1+e^{-z})^{-1}$ for the second layer. A subset of the pre-computed spectral grid (that is $15\,000$ synthetic spectra) is used to train the ANN, whereby the weights and the offsets are adjusted to the optimal values. This subset is referred to as a \textit{training set}. We train the neural networks by minimising the $L^2$ loss. In other words, we compute a minimal sum of the Euclidean distances between the target ab-initio flux from the training set and the flux predicted by the model at each wavelength point. We use cross-validation with the remaining set of $5000$ spectra, which are referred to as a \textit{cross-validation set} to prevent over-fitting. This requires optimal values of the ANN to decrease $L^2$ loss also for the \textit{cross-validation set}, which is not directly used during training. Together, the ANN layers act like a function that predicts a flux spectrum for a set of given labels. The main difference of the current implementation of \textit{Payne} with respect to the one in \citet{ting2018} is that we use only one ANN to represent the full stellar spectrum. In our realisation\footnote{as it is now implemented in the Github version: \url{https://github.com/tingyuansen/The_Payne}.} an ANN can exploit information from the adjacent pixels, 
while previously each individual pixel was trained separately.
A synthetic spectrum is generated at arbitrary points in stellar parameter space within the domain of the training grid and is compared to the observed spectrum. A standard $\chi^2$ minimisation is used to compute the likelihood of the fit and, hence, to find the stellar parameters that best characterise the observed spectrum. We also allow for a small Doppler shift, $\pm~2~\kms$, on top of the RV from cross-correlation, to optimise the spectral fit.

The continuum normalisation of the observed spectra is performed during the $\chi^2$ minimisation. We search for the coefficients of a linear combination of the first ten Chebyshev polynomials, which represents a function that fits the shape of the continuum, using the full observed spectrum. A synthetic spectrum is then multiplied with this function.

In total, for each observed spectrum, we optimise 19 free parameters: one Doppler shift, eight spectral labels and ten coefficients of Chebyshev polynomials. The abundances of individual elements are derived simultaneously with other stellar parameters via the full spectral fitting process. We also employed the classical method of fitting separately each spectral line using line masks. However, this method delivers less precise abundances, as gauged by the star-to-star scatter, hence, we do not use the line masks in the final abundance analysis.

Following the result in \citet{Bergemann2011} which strongly recommended to use only Ti II lines in abundance analysis, we masked out all Ti I lines. We note, however, that we did not include NLTE calculations for Ti II, as the NLTE effects on this ion are very small in the metallicity regime of our sample \citep{Bergemann2011}. Hence, the difference between our LTE and NLTE Ti abundance reflects only an indirect effect of NLTE on stellar parameters. 
\subsection{Internal accuracy of the method} 
\label{cvtest}
We verify the internal accuracy of the method by subjecting it to tests similar to those employed by \citet{ting2018}.

First, we compare the interpolated synthetic spectra to the original models from the cross-validation sample. In this case we explore how well \textit{the Payne} can generate new spectrum. The median interpolation error of the flux across $5000$ models is $\leq 10^{-3}$, that is, within $0.1$\%. We also find that larger errors occur for cooler stars, because there are many more spectral features. This result suggests that interpolation is more accurate than the typical $\snr$ of observed spectrum.

Second, we test how well we can recover original labels from the model, through $\chi^2$ minimisation. In this case we apply random Doppler shift, multiply the model spectrum by a random combination of the first ten Chebyshev polynomials, that represent the continuum level and add noise. Such a modified model serves as a fair representation of a real observed spectrum. The tests are performed for the noiseless models and the models degraded to a $\snr$ of 90 \AA$^{-1}$ and 224 \AA$^{-1}$. This range of $\snr$ brackets the typical values of the observed HR10 spectra, with the majority of the spectra sampling the $\snr$ range of 150-200 \AA$^{-1}$. The typical $\snr$ of the spectra of the benchmark stars is $\sim 200$ \AA$^{-1}$.

Table~\ref{tab:interr} presents the average differences between the input and the output stellar parameters for the cross-validation sample. The scatter is represented by one standard deviation. To facilitate the analysis, we group the results into three metallicity bins.

The results for the noiseless models with [Fe/H] in the range from $-1.6$ to $0.5$ dex suggest high internal accuracy of the method. For the lower-metallicity models, there is a small bias and a larger dispersion in the residuals, because we have less spectral information in this regime. The bias is also marginal for the high-$\snr$ spectra with $\snr = 224$ \AA$^{-1}$, although the scatter in the output is increased compared to the noiseless models. Our analysis of the noisy models, $\snr = 90$ \AA$^{-1}$, yields acceptable results for the metal-rich and moderately metal-poor stars with $\feh \gtrapprox -1.6$~dex. On the other hand, the most metal-poor noisy spectra are not fitted well. Despite a modest bias in $\teff$, the dispersion of $\log g$ and the abundance ratios is very large and may require a different approach to obtain high-precision abundances in this regime. According to this test, good Mn abundances (better than $\sim 0.1$~dex) can be derived only for metal-rich stars.   

These tests illustrate only the internal accuracy of \textit{the Payne} model reconstruction and, hence, set the minimum uncertainty on the parameters determined by our method, regardless of the training sample, its physical properties and completeness. The analysis of observed data may result in a larger uncertainty, as various other effects, such as the physical complexity of the model atmospheres and synthetic spectra and properties of the observed data (data reduction effects etc.), will contribute to the total uncertainties. We test this in the next section by analysing the Gaia-ESO benchmark stars.
%
%
\begin{table*}[tp]\small
	\centering
	
	\caption{Internal errors of label's recovery by \textit{the Payne} see Section~\protect\ref{cvtest} for details.}
	\label{tab:interr}
	
	\begin{tabular}{lccccccccc} 
		\hline
		\hline
		$\snr$ &$\feh$&$\Delta\teff$ &$\Delta\logg$ &$\Delta\Vmic$ &$\Delta\Vbrd$&$\Delta\feh$ &$\Delta\mgfe$ &$\Delta\tife$&$\Delta\mnfe$\\
		\AA$^{-1}$&dex&~1000~K&dex &~$\kms$& ~10~$\kms$ &dex&dex&dex&dex\\
		\hline
90&-2.6:-1.6&0.00$\pm$0.27&-0.05$\pm$0.56 &-0.10$\pm$0.77 &-0.01$\pm$0.29 &-0.00$\pm$0.18 &0.01$\pm$0.17 &-0.02$\pm$0.34&-0.01$\pm$0.64 \\
&-1.6:-0.6&0.01$\pm$0.12&0.00$\pm$0.21 &0.01$\pm$0.26 &-0.00$\pm$0.09 &0.01$\pm$0.07 &0.00$\pm$0.10 &-0.01$\pm$0.13&-0.01$\pm$0.37 \\
&-0.6:0.5&0.01$\pm$0.07&0.00$\pm$0.12 &0.00$\pm$0.10 &-0.00$\pm$0.05 &0.01$\pm$0.06 &-0.00$\pm$0.09 &-0.00$\pm$0.07&-0.00$\pm$0.11 \\
\hline
224&-2.6:-1.6&-0.00$\pm$0.11&-0.04$\pm$0.25 &-0.00$\pm$0.41 &0.01$\pm$0.13 &-0.00$\pm$0.07 &0.01$\pm$0.07 &-0.01$\pm$0.17&-0.03$\pm$0.55 \\
&-1.6:-0.6&-0.00$\pm$0.05&-0.00$\pm$0.08 &-0.00$\pm$0.09 &-0.00$\pm$0.04 &-0.00$\pm$0.03 &0.00$\pm$0.05 &-0.00$\pm$0.05&-0.03$\pm$0.25 \\
&-0.6:0.5&0.00$\pm$0.03&0.00$\pm$0.06 &0.00$\pm$0.05 &-0.00$\pm$0.02 &0.00$\pm$0.04 &0.00$\pm$0.04 &0.00$\pm$0.03&0.00$\pm$0.05 \\
\hline
no&-2.6:-1.6&-0.00$\pm$0.02&-0.01$\pm$0.06 &-0.01$\pm$0.15 &0.00$\pm$0.02 &-0.00$\pm$0.02 &-0.00$\pm$0.03 &-0.00$\pm$0.04&-0.01$\pm$0.21 \\
noise&-1.6:-0.6&-0.00$\pm$0.01&0.00$\pm$0.03 &0.00$\pm$0.03 &0.00$\pm$0.01 &0.00$\pm$0.01 &-0.00$\pm$0.02 &0.00$\pm$0.02&0.00$\pm$0.05 \\
&-0.6:0.5&0.00$\pm$0.01&0.00$\pm$0.05 &0.00$\pm$0.03 &0.00$\pm$0.01 &0.00$\pm$0.05 &-0.00$\pm$0.04 &0.00$\pm$0.01&0.00$\pm$0.03 \\
\hline
\end{tabular}
\end{table*}

%
%
\section{Results}\label{Results}
\subsection{Gaia-ESO benchmark stars}
\begin{figure*}
\includegraphics[width=\textwidth]{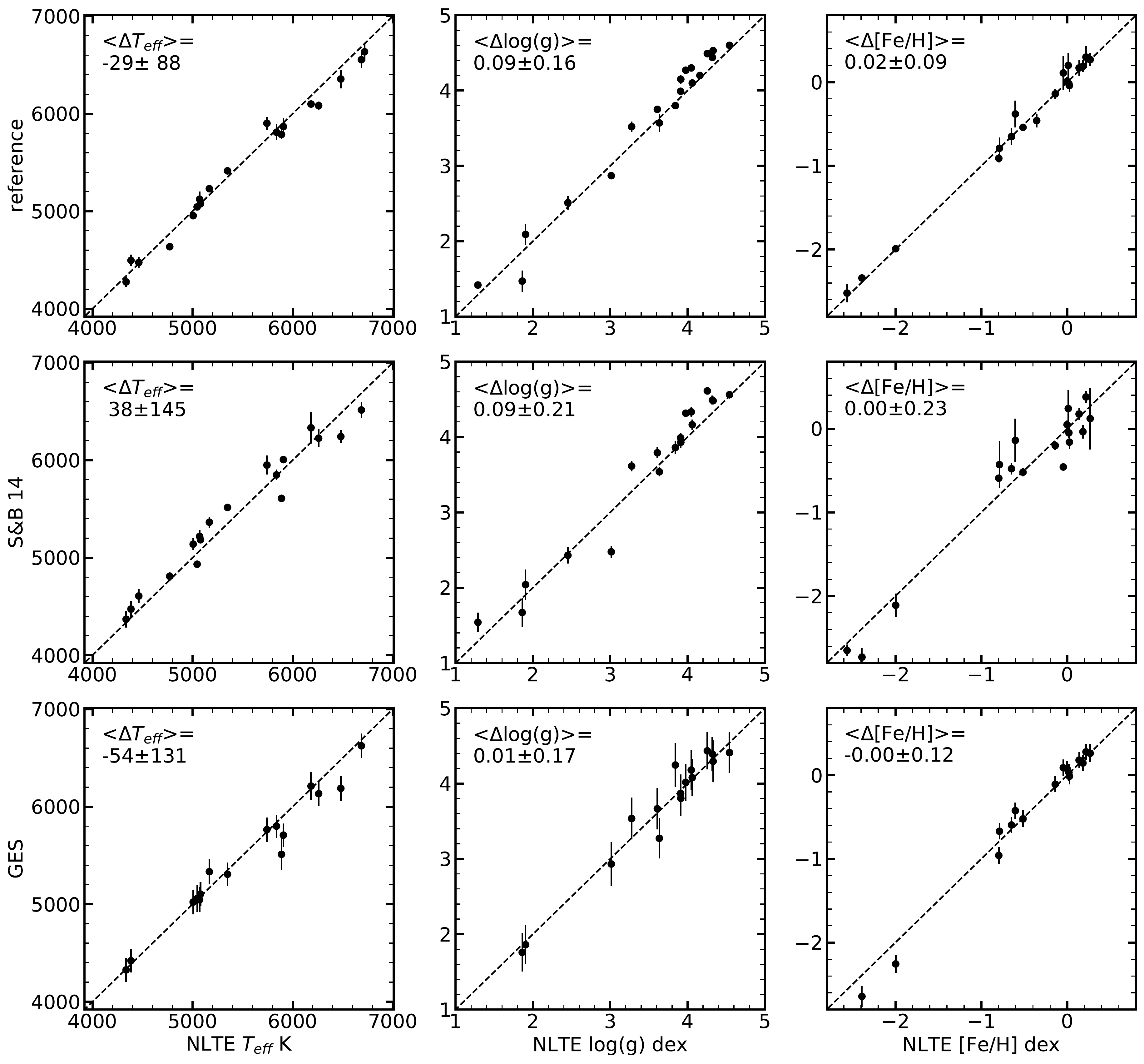}
\caption[]{Our NLTE spectroscopic estimates for the benchmark stars compared with the literature. The top panels shows the reference stellar parameters and their uncertainties from \citet{Jofre2015,Karovicova2018,amarsi2016}. In the middle and bottom panels, we show our values against the results from \citet{Schoenrich2014} and GES catalogue \protect\citet{Smiljanic2014a}, respectively. The mean offset and scatter are given in the legend of each plot. }
\label{fig:2} 
\end{figure*}

\begin{figure}
\includegraphics[width=\columnwidth]{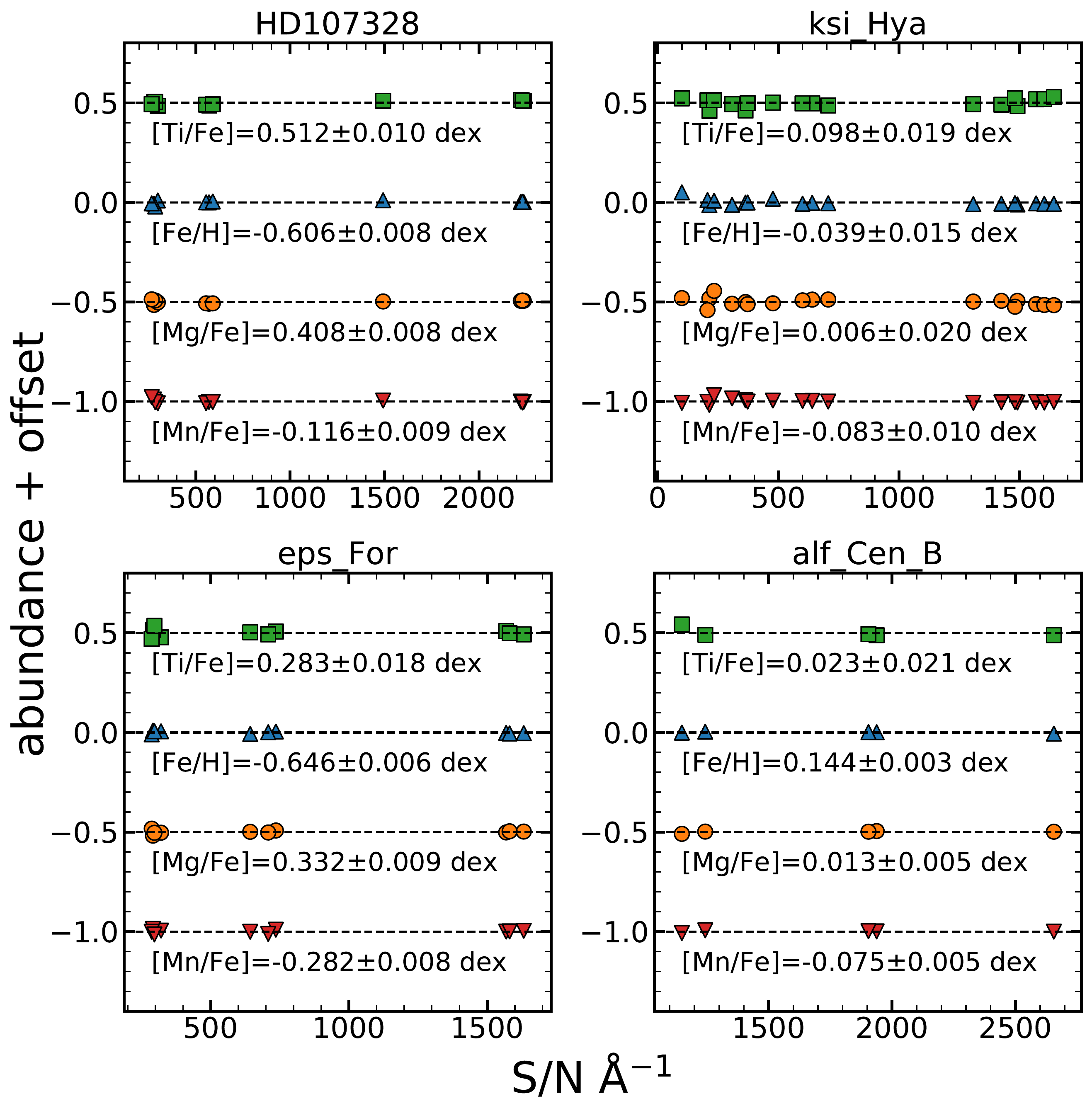}
\caption[]{NLTE elemental abundances derived from the spectra taken at different exposure times. Abundances determined at $\snr=200$~\AA$^{-1}$ are just as precise as those at $\snr>2000$~\AA$^{-1}$, see section~\protect\ref{gaiabenchmark} for details.}
\label{fig:3} 
\end{figure}

\label{gaiabenchmark}
Our results for the Gaia-ESO benchmark stars are shown in Fig.~\ref{fig:2} and Fig.~\ref{fig:3}.

Fig. \ref{fig:2} compares our NLTE stellar parameters with the values from \citet{Jofre2015}, \citet{Schoenrich2014}, and with the Gaia-ESO DR3 catalogue (GES) \citep{Smiljanic2014a}. In \citet{Jofre2015}, $\teff$ estimates were determined from photometry and interferometry, $\logg$ from parallaxes and astroseismology. $\feh$ estimates were obtained from the NLTE analysis of Fe lines in the high-resolution spectra taken with the UVES, NARVAL and HARPS spectrographs \citep{blanco-cuaresma2014}. In order to be consistent with our reference solar $\feh$ scale, we subtracted 0.05 dex from Jofre and GES metallicities, as they are based on the \protect\citet{Grevesse2007} metallicity scale (logA(Fe)$_\odot=7.45$~dex). Likewise, we subtracted 0.03 dex from \citet{amarsi2016} metallicities, as they employ logA(Fe)$_\odot=7.47$~dex. %
The estimates of stellar parameters in \citet{Schoenrich2014} are derived using a full Bayesian approach by solving for the posterior in a multi-dimensional parameter space, including photometry, high-resolution spectra, parallaxes, and evolutionary constraints. The estimates of stellar parameters in the Gaia-ESO DR3 catalogue rely on the high-resolution (UVES at VLT) spectroscopy only.

Fig. \ref{fig:2} suggests that the agreement of our NLTE results with the literature studies is very good. The differences with \citet{Jofre2015} are of the order  -29$\pm$88~K in $\teff$, 0.09$\pm$0.16 dex in $\logg$ and 0.02$\pm$0.09 dex in $\feh$ across the full parameter space, and they also compare favourably with the results obtained by \citet{Schoenrich2014} and reported in the Gaia-ESO DR3 catalogue. Results for individual stars are listed in Table~\ref{tab:gbs}. 
The scatter is slightly larger for the metal-poor stars. This could be the consequence of the limited coverage of the training set. In particular, the two very metal-poor evolved stars HD~122563 and HD~140283 are located next to the low-metallicity edge of our training grid. Since the Gaia-ESO benchmark star sample contains only three stars with $\feh< -1$, no reliable statistics can be drawn on the success of our approach in this regime of stellar parameters. Also the sample of RGB stars is very small and contains  only five objects with $\logg < 3$~dex. We address the performance of our method for low-gravity stars in the next section, by analysing a set of open and globular clusters that cover a large metallicity range, $-2.3 \lesssim \feh \lesssim -0.1$ dex, and provide a better sampling on the RGB.%

Fig. \ref{fig:3} illustrates the performance of our method for the spectra taken at different exposure times. We have chosen four stars representative of our calibration sample: HD~107328 - a moderately metal-poor giant ($\teff = 4384$ K, $\logg = 1.90$~dex, and $\feh_{\rm NLTE} = -0.60$ dex), $\xi~Hya$ - a metal-rich subgiant ($\teff = 5045$ K, $\logg = 3.01$~dex, and $\feh_{\rm NLTE} = -0.05$ dex), $\epsilon~For$ - a moderately metal-poor subgiant ($\teff = 5070$ K, $\logg = 3.28$~dex, and $\feh_{\rm NLTE} = -0.65$ dex), and $\alpha~Cen~B$ - a metal-rich dwarf ($\teff = 5167$ K, $\logg = 4.33$~dex, and $\feh_{\rm NLTE} = 0.14$ dex). These stars have been observed with different exposure times, corresponding to the $\snr$  ratios of $100$ to $2500$ \AA$^{-1}$ that allows us to validate the differential precision of the adopted model. We do not detect any evidence of a systematic bias  that depends on the data quality. In particular, the mean difference (taken as one standard deviation) between abundances of Fe, Mg, and Ti obtained from the $\snr \sim 100$ \AA$^{-1}$ spectra and those obtained from the highest-quality data ($\snr\sim2000$ \AA$^{-1}$) is not larger than $0.02$ dex for any of these stars, and is less than $0.01$ dex for the majority. We hence conclude that our results are not very sensitive to the quality of the observed data for a wide range of $\snr$ ratios.
\begin{figure*}
\includegraphics[width=0.99\textwidth]{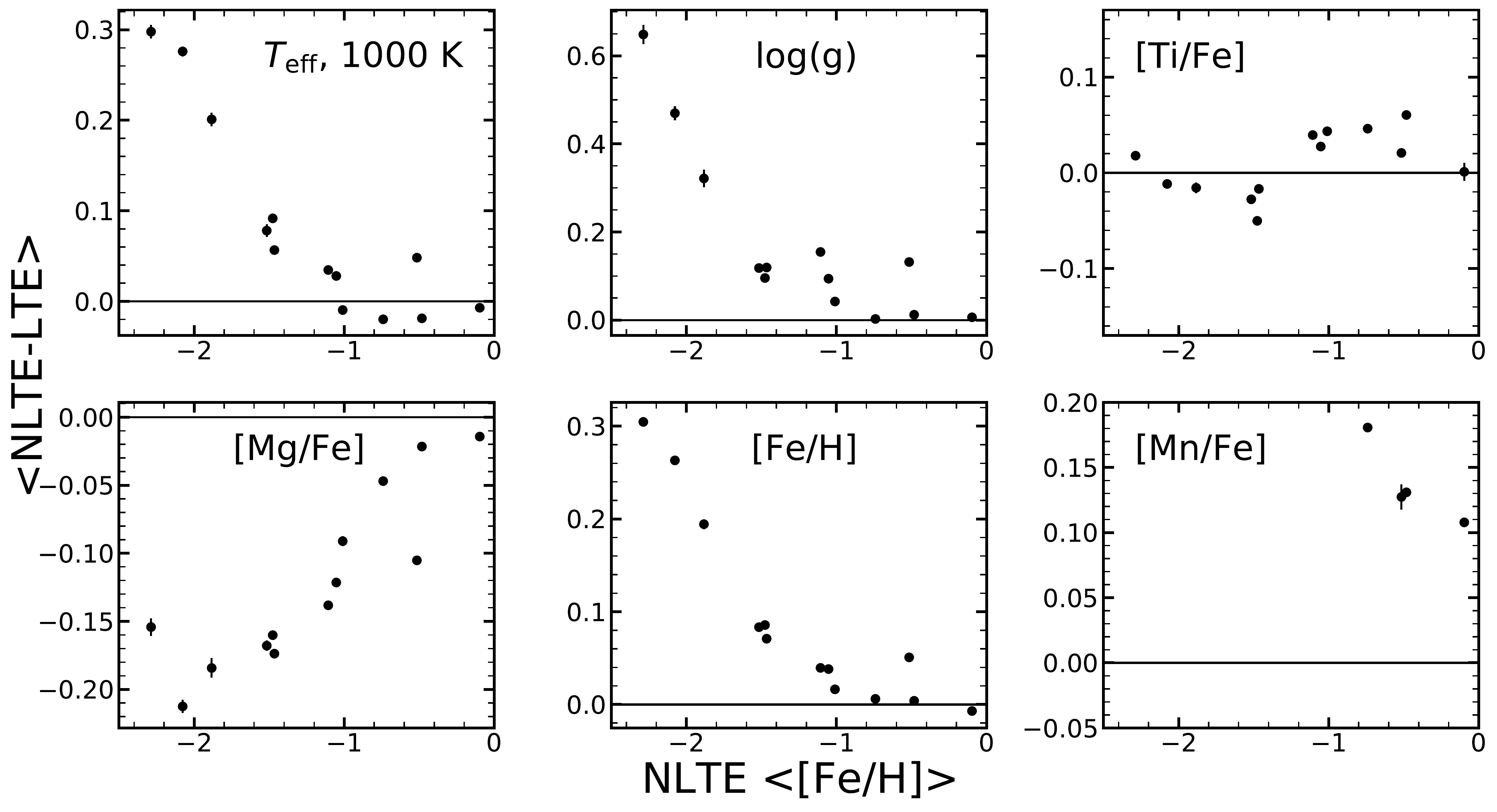}
\caption{Mean differences between NLTE and LTE parameters for stars within each cluster against NLTE metallicity. For $\mnfe$ only clusters with $\feh>-1$ dex are shown. See section \protect\ref{isochrones} for details.}
\label{fig:4} 
\end{figure*}
%
%
\subsection{Open and globular clusters}
\subsubsection{Sample selection}
Our dataset includes two open clusters and $11$ globular clusters. The cluster members are chosen using the central coordinates and the RV estimates from the SIMBAD\footnote{\url{http://simbad.u-strasbg.fr/simbad/}} database listed in Table~\ref{tab:gcinfo}. We select only stars with an RV within 5~$\kms$ from the cluster median\footnote{The median is used because it is less sensitive to outliers.}, for the open clusters. For the globular clusters, we assume a 1$\sigma$ RV dispersion and the central values from \citet{Pancino2017}. We also apply a 2$\sigma$ clipping around the median in metallicity, and employ proper motions from Gaia DR2 \citep{gdr2} to exclude stars outside the 2$\sigma$ range from the median proper motion of each cluster.
It is common to use distances to compute astrometric gravities \citep[e.g.][]{ruchti2013}. However, the majority of clusters in our sample are located at heliocentric distances $d_\odot$ of $> 2 $ kpc, where parallaxes are very uncertain. Besides, poorly constrained differential extinction in some clusters limits the applicability of standard relations, to derive log(g) from distances and photometric magnitudes. We, hence, refrain from using the GDR2 parallaxes to compute surface gravities. Instead, we compare our results with the isochrones computed using our estimates of metallicities and the ages adopted from literature studies, in particular, from  \citet{kruijssen2018} for GCs and from the WEBDA database\footnote{\url{https://www.univie.ac.at/webda/}} for open clusters. For most clusters, the ages are derived from the colour-magnitude diagram turn-off (TO) or horizontal branch (HB) fits. Hence, also this comparison can be performed only with the caveat that the TO/HB ages are not a fundamental reference, but are model-dependent and may not be fully unbiased.
\begin{figure*}
\includegraphics[width=\textwidth]{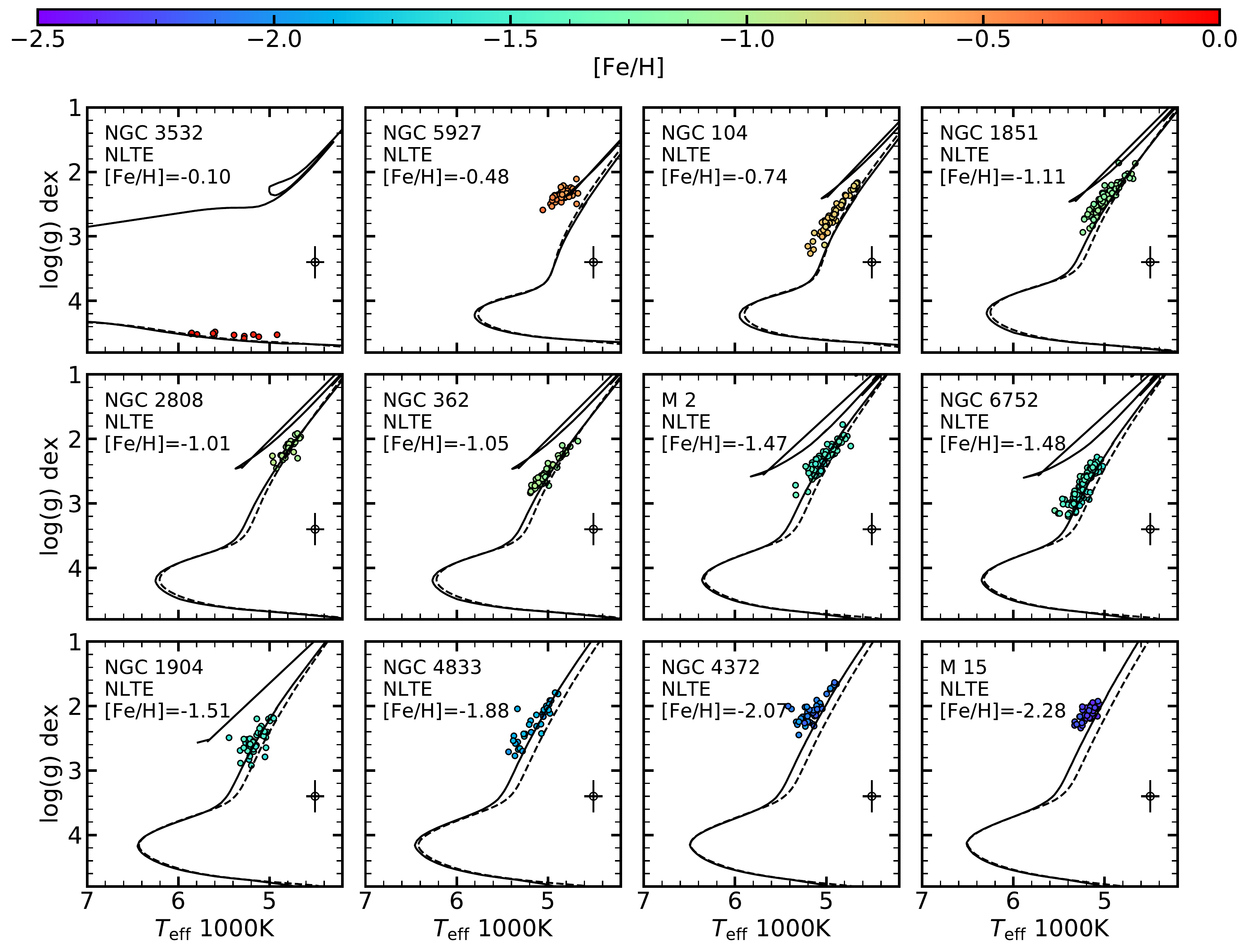}
\caption{NLTE spectroscopic parameters compared with the PARSEC (solid line) and Victoria-Regina (dashed line) isochrones \citep{marigo2017,VR}. The isochrones were computed using literature ages and $\feh$ from our NLTE analysis. The colour of the points indicates their [Fe/H].  We note the different target selection for NGC 5927, where the observed stellar sample contains mostly red clump stars. }
\label{fig:5} 
\end{figure*}

\begin{figure*}
\includegraphics[width=\textwidth]{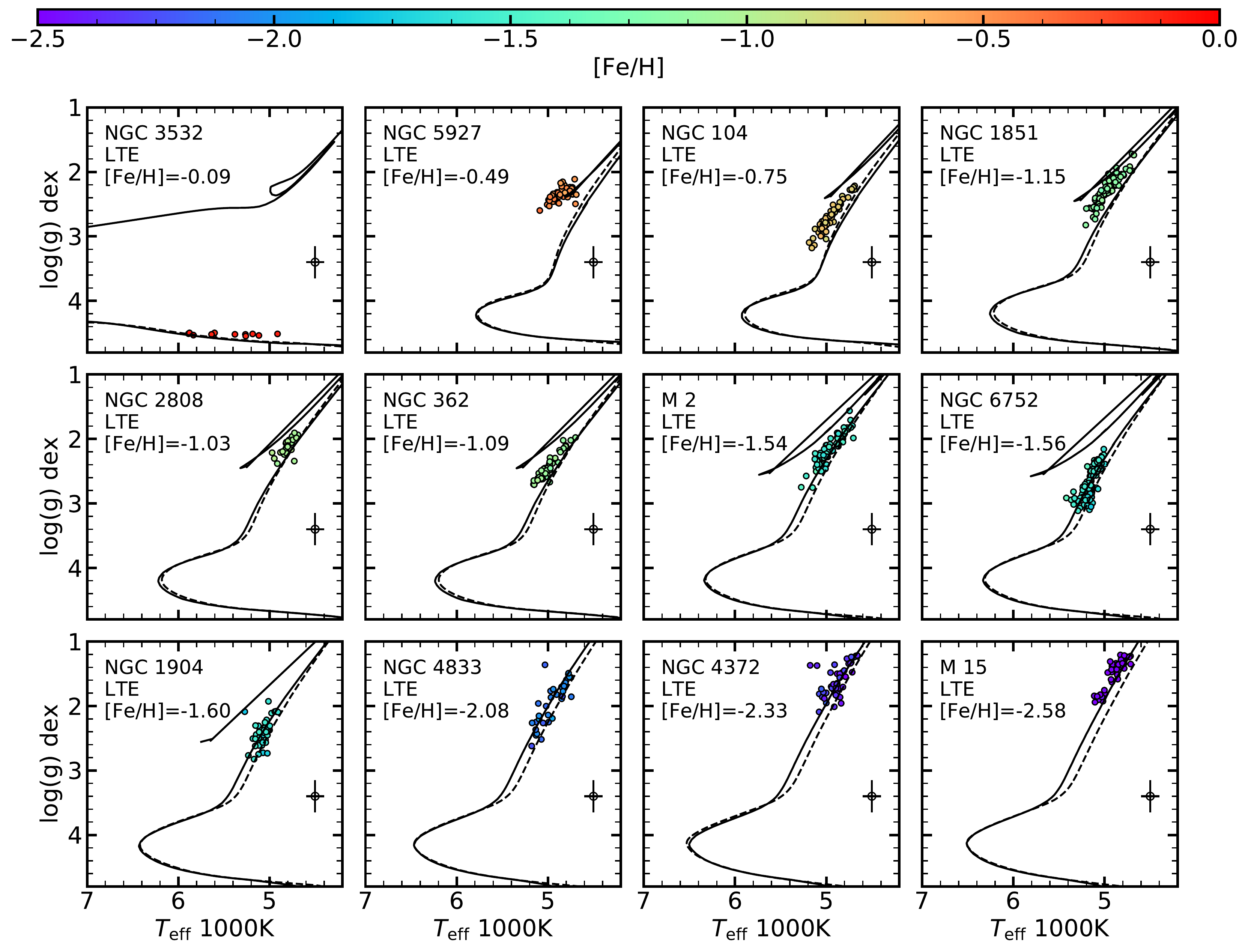}
\caption{LTE spectroscopic parameters compared with the PARSEC (solid line) and Victoria-Regina (dashed line) isochrones \citep{marigo2017,VR}.  The isochrones were computed using literature ages and $\feh$ from our LTE analysis. The colour of the points indicates their [Fe/H]. We note the different target selection for NGC 5927, where the observed stellar sample contains mostly red clump stars.}
\label{fig:6} 
\end{figure*}
%
%
%
\subsubsection{Stellar parameters and comparison with the isochrones}
\label{isochrones}
The majority of the globular clusters are distant and are represented by RGB stars in our sample. Main-sequence stars are observed only in the nearby metal-rich open cluster NGC 3532. Hence, in what follows, the discussion will mainly focus on the RGB population across a wide range of metallicities, from $-0.5$ (NGC 5927) to $-2.3$~dex (M 15).
\par
In Fig. \ref{fig:4}, we compare NLTE and LTE stellar parameters as a function of NLTE metallicity. Since most stars, within a cluster, are in the same evolutionary stage (lower or upper RGB), we have chosen to show only the mean  NLTE-LTE differences, averaged over all stars in a given cluster. This is sufficient to illustrate the key result: the differences between NLTE and LTE measurements of $\teff$, $\log g$ and $\feh$ vary in lockstep with metallicity. This reflects the NLTE effects in the formation of the Fe I and Ti I spectral lines, which are ubiquitous in HR10. It is furthermore important, although not unexpected, that below $\feh \sim -1$~dex the changes are nearly linear, consistent with our earlier theoretical estimates \citep{lind2012} and with the analysis of the metal-poor field stars in the Milky Way \citep{ruchti2013}. The NLTE effect is most striking at $\feh \lesssim -2$, where we find differences of $\sim 300$ K in $\teff$, $\sim 0.6$ dex in $\log g$ and $\sim 0.3$ dex in \feh. The [Mg$/$Fe] ratios tend to be lower in NLTE, this reflects negative NLTE abundance corrections for the only Mg line in HR10 (Mg I 5528 \AA), which is consistent with earlier studies \citep{osorio2015, bergemann2017b}. The upturn in [Mg$/$Fe] at $\feh\sim -2$~dex is real and it is caused by the change of the dominant NLTE effect at this metallicity. At higher [Fe/H], strong line scattering and photon loss, and, hence, the deviations of the source function from the Planck function, play an important role in the statistical equilibrium of the ion. However, in the metal-poor models, [Fe/H] $\lesssim -2$~dex, it is the over-ionisation driven by a hard UV radiation field that acts on the line opacity and thereby counteracts the NLTE effects on the source function. We have masked out all Ti I lines (see Section~\ref{payne}), so differences in $\tife$ are small $\lesssim0.06$~dex and represent indirect NLTE effects on other stellar parameters. The difference in $\mnfe$ is shown only for a few metal-rich clusters, and it is increasing to lower $\feh$. 
\par
Fig. \ref{fig:5} and Fig. \ref{fig:6} show our NLTE/LTE results, respectively for the $12$ clusters in the $\teff$ - $\log g$ plane. We also overlay the PARSEC \citep{marigo2017} and Victoria-Regina \citep[][hereafter, VR]{VR} isochrones to facilitate the analysis of the evolutionary stages probed by the stellar sample. The VR isochrones assume the He abundance of $Y=0.26$ and an  $\alpha$-enhancement, as given by our measurements of $\mgfe$. The PARSEC isochrones are computed using an effective metallicity (Aldo Serenelly, priv. comm.)
\begin{equation}
Z=Z_{0}(0.659 f_{\alpha}+ 0.341),
\end{equation}
where $Z_{0}=10^{\feh}$ and $f_{\alpha}=10^{\mgfe}$. The error of the spectroscopic estimates is shown in the inset and it represents the typical uncertainty of our analysis ($\Delta (\teff) = 150$ K and $\Delta (\logg) = 0.3$ dex based on analysis of Gaia-ESO benchmark stars). The star-to-star scatter in the $\teff$-$\logg$ plane is very small, and, within the uncertainties, consistent with the isochrones.
\par
Surprisingly, both NLTE and LTE spectroscopic parameters agree well with the isochrones computed for the corresponding [Fe/H], despite the large differences between NLTE and LTE parameters ($\teff$, $\log g$, $\feh$, and $\mgfe$) especially at low metallicity. This would appear counter-intuitive, at first glance, given the large offsets demonstrated in Fig. \ref{fig:4}. However, this effect is, in fact, simply a result of the complex correlations in stellar parameters \citep[as also extensively discussed in][]{ruchti2013}: NLTE effects in the over-ionisation - dominated species (such as Fe I, Ti I) significantly change the excitation and ionisation balance, such that the theoretical spectral lines tend to be weaker and a higher abundance would be inferred by comparing them to the observed spectra. Consequently, larger estimates of $\teff$, $\log g$, and $\feh$ are expected from the NLTE modelling compared to LTE \citep[see also][]{lind2012}. The difference between NLTE and LTE \feh~estimates is exactly the offset needed to match the higher (lower) $\teff$ and higher (lower) $\log g$ to the corresponding isochrone computed for the NLTE (LTE) metallicity and $\alpha$-enhancement. This suggests that even large systematic errors in spectroscopic estimates may remain undetected in the $\teff - \log g$ plane, when spectroscopic values are gauged by comparing them with isochrones.
\par
\begin{figure*}
    \centering
    \includegraphics[width=\textwidth]{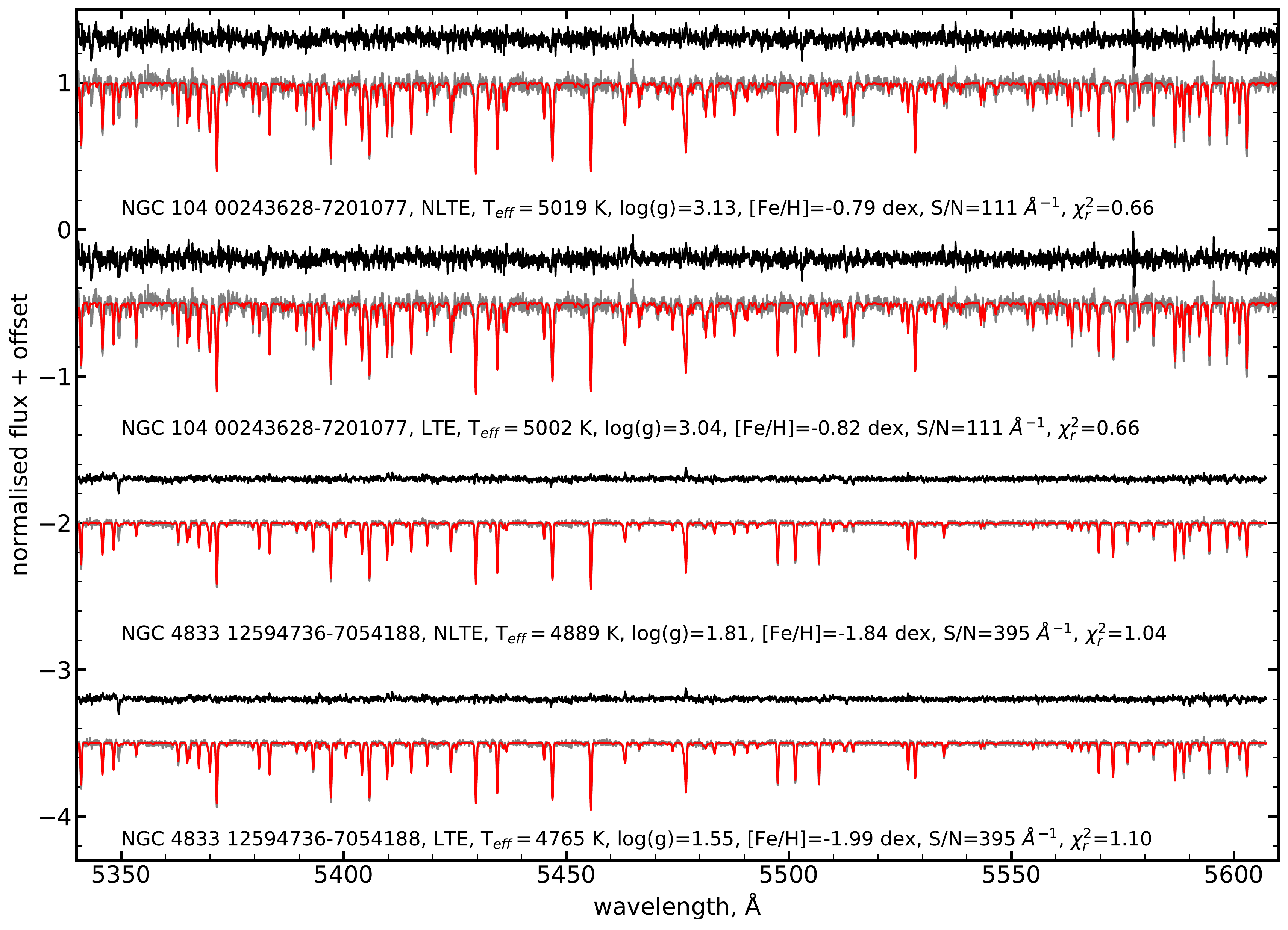}
    \caption{NLTE and LTE spectral fits for two stars from our cluster sample. The best-fit spectra are shown as red lines. The observed spectra are depicted with grey lines, while the fit residuals are shown with black lines. See inset for the derived $\teff, \logg, \feh$, reduced $\chi^2$, and the $\snr$ of the observed spectrum.}
    \label{fig:fit}
\end{figure*}

In Fig.~\ref{fig:fit} we show examples of spectral fits for two stars randomly selected from our sample of clusters. Both LTE and NLTE model spectra match the observed ones very well, having similar $\chi^2_r$, while the fit residuals mostly show noise and data reduction artefacts.  

\par
Our LTE and NLTE results show a slight tendency towards a hotter $\teff$ scale, which may appear more consistent with the PARSEC models. However, it might be premature to draw more specific conclusions on this matter, as we are aware of the imperfections of the stellar atmosphere and spectral model grids, such as an approximate treatment of convection as well as calibrations that are employed in the stellar evolution models \citep[e.g.][]{fu2018}). At this stage, it appears to be sufficient to emphasise that our spectroscopic results are internally consistent, and allow predictive statements to be made on the astrophysical significance of the similarities and/or differences of chemical abundance patterns in the clusters.
\begin{figure*}
\includegraphics[width=\textwidth]{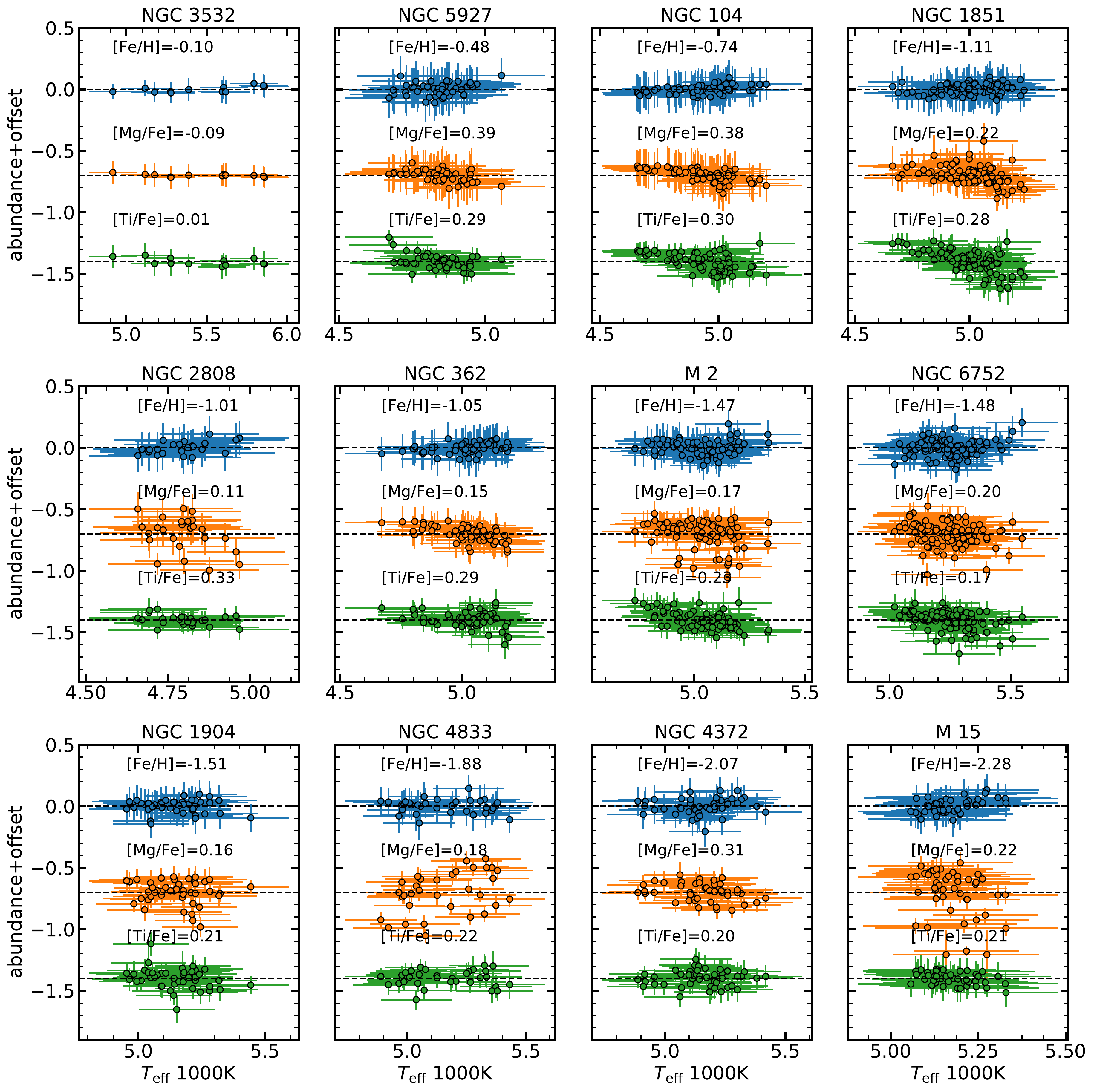}
\caption{NLTE abundances as a function of $\teff$ for all cluster stars. The average is shown for all elements. The scatter in $\mgfe$ is much larger than in $\feh$ and $\tife$, and it is typically attributed to multiple episodes of star formation and self-enrichment \protect\citep[see the recent review by ][and references therein]{Bastian2018}. See Section ~\protect\ref{trends} for details. }
\label{fig:8} 
\end{figure*}
%
%
\subsubsection{Error estimates}
\label{systerr}
To explore the sensitivity of the abundances to the uncertainties in stellar parameters, we use a method similar to the one employed in \citet{Bergemann2017a}. The standard errors are estimated by comparison with the independent stellar parameters for the benchmark stars (Section~\ref{gaiabenchmark}). These are $\pm\Delta \teff=150$~K, $\pm\Delta \logg=0.3$~ dex and $\pm\Delta \feh=0.1$~dex. For $\Vmic$, we use the  uncertainty of $\pm~0.2~\kms$. We perturb one parameter at a time by its standard error, and re-determine the abundance of an element, while keeping the parameter fixed during the $\chi^2$ optimisation. We then compare the resulting abundance with the estimate obtained from the full solution, when all labels are solved for simultaneously. Table~\ref{tab:sist} presents the resulting uncertainties for five stars representative of the sample. These differences are added in quadrature and are used as a measure of the systematic error of abundances --$\Delta X$. The systematic errors derived using this procedure are typically within $0.10$ to $0.15$ dex (Table~\ref{tab:sist}).

The test of internal accuracy suggests  (Section~\ref{cvtest}) that we cannot have derived robust Mn abundances for much of the parameter space, because Mn lines in the HR10 spectra are weak in the metal-poor regime. Hence, the mean [Mn/Fe] ratios are only provided for the two metal-rich clusters NGC 3532 and NGC 5927.
\subsubsection{Abundance spreads in clusters}
\label{trends}
\begin{figure}
    \centering
    \includegraphics[width=\columnwidth]{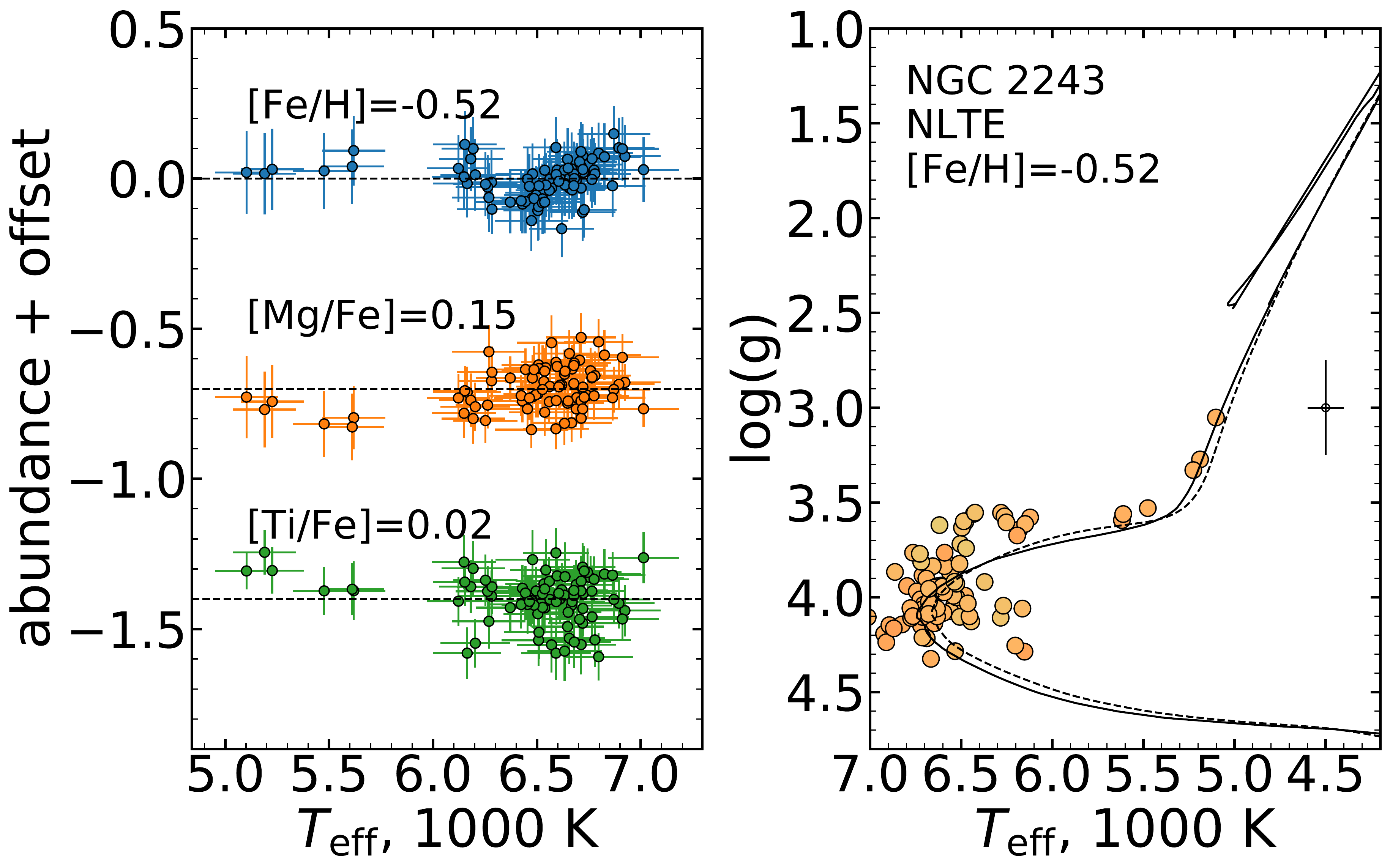}
    \caption{Abundances as a function of $\teff$ and the $\teff$-$\logg$ diagram for the open cluster NGC~2243. All values are our NLTE results. The isochrones were computed for the age of $3.8$ Gyr from \protect\citet{Anthony-Twarog2005} and $\feh_{\rm NLTE}=-0.52$~dex.}
    \label{fig:9}
\end{figure}

Fig. \ref{fig:8} shows the [Fe/H], [Ti/Fe], and [Mg/Fe] abundance estimates in stars of OCs and GCs against stellar $\teff$. The uncertainties represent the systematic errors computed as described in see Section~\ref{systerr}. The open cluster NGC~2243 is shown separately in Fig.~\ref{fig:9} 
as it shows signatures of atomic diffusion. Of a particular interest is the dip of [Fe/H] at the cluster TO ($\teff \sim 6400$~K), which is qualitatively consistent with the predictions of stellar evolution models, which include radiative acceleration and gravitational settling \citep[e.g.][]{deal2018}. We leave a detailed exploration of this effect for our future study.  

Whereas prominent systematic biases appear to be absent for most clusters, there is some evidence for a small anti-correlation of $\mgfe$ and/or $\tife$ values with $\teff$, for the moderately metal-poor clusters NGC~1851, NGC 362, M2, and NGC 6752. These clusters also show a somewhat tilted distribution of stars relative to the isochrones in the $\teff - \log g$ plane (Fig.\ref{fig:5},\ref{fig:6}) suggesting that the origin of the trends is likely in the spectral models/method, employed in this work. Currently we have no straightforward solution for this effect.

 The average abundance of a cluster <$X$> and internal dispersion $\sigma_X$ are computed using maximum likelihood (ML) approach \citep{Walker2006,Piatti2018}, where we take into account the individual abundance uncertainties $\Delta X$ of each star. We numerically maximise the logarithm of the likelihood $L$, given as:
\begin{equation}
    \label{likelihood}
    \ln{L}=-\frac{1}{2} \sum_i^N \ln(\Delta X_i^2+\sigma_X^2) - \frac{1}{2} \sum_i^N \frac {(X_i-<X>)^2}{\Delta X_i^2+\sigma_X^2} -\frac{N}{2}\ln{2\pi}
\end{equation}
where \textit{N} is the number of stars in a cluster and \textit{X} refers to one of $\feh,~\mgfe,~\tife$ and  $\mnfe$. The errors of the mean and dispersion are computed from the respective covariance matrices \citep{Walker2006}.  

We find that all clusters are homogeneous in $\feh$ and $\tife$ at an uncertainty level of $0.03$~dex. Four clusters (M~15, M2, NGC~4833, NGC~2808) show a larger scatter in $\mgfe$ at the level of $0.07$~dex or greater. Modest internal dispersions $\sigma_{\mgfe} \sim 0.04$~dex are detected in NGC~1904 and NGC~6572. 

Spreads in light element abundances, including Mg, have already been reported for a number of clusters, including NGC 2808 \citep{carretta2015}, M2 \citep{yong2015}, NGC 4833 \citep{Carretta2014a} and M15 \citep{carretta2009}. These spreads are typically attributed to multiple episodes of star formation and self-enrichment \citep[see the recent review by ][and references therein]{Bastian2018}. 

 The estimated internal dispersions are summarised in Table~\ref{tab:newnlte1}. In the following, to be consistent with the literature, we will focus on the {\em observed} intra-cluster dispersion, instead of the ML estimated internal dispersion. We note that these two are not the same as the latter probes the {\em intrinsic} dispersion that is not accounted for by the measurement uncertainties, while the former includes both.

\begin{table*}[tp]\small
    \centering
    \caption{Mean clusters abundances (ML estimate) with observed intra-cluster spread (standard deviation) and mean systematic error (err).}
    \begin{tabular}{lccccccc}
\hline
Cluster & \#stars  & $\feh_{\rm NLTE}$~dex & $\feh_{\rm LTE}$~dex & $\mgfe_{\rm NLTE}$~dex & $\mgfe_{\rm LTE}$~dex & $\tife_{\rm NLTE}$~dex & $\tife_{\rm LTE}$~dex\\
 &N&avg std <err>&avg std <err>&avg std <err>&avg std <err>&avg std <err>&avg std <err>\\
\hline
NGC 3532 &12&-0.10 0.02 0.10&          -0.09 0.03 0.11&          -0.09 0.01 0.09&          -0.07 0.01 0.12&          0.01 0.03 0.10&          0.01 0.03 0.11\\
NGC 5927 &47&-0.48 0.05 0.16&          -0.49 0.05 0.16&          0.39 0.04 0.07&          0.41 0.05 0.07&          0.29 0.06 0.07&          0.23 0.05 0.07\\
NGC 2243 &84&-0.52 0.06 0.08&          -0.57 0.07 0.11&          0.15 0.07 0.08&          0.26 0.09 0.09&          0.02 0.08 0.09&          0.01 0.09 0.10\\
NGC 104 &68&-0.74 0.03 0.15&          -0.75 0.03 0.17&          0.38 0.05 0.08&          0.42 0.04 0.08&          0.30 0.07 0.08&          0.26 0.07 0.08\\
NGC 1851 &88&-1.11 0.04 0.14&          -1.15 0.04 0.15&          0.22 0.08 0.10&          0.36 0.08 0.08&          0.28 0.09 0.11&          0.24 0.07 0.10\\
NGC 2808 &25&-1.01 0.05 0.14&          -1.03 0.05 0.15&          0.11 0.14 0.08&          0.22 0.15 0.06&          0.33 0.04 0.07&          0.30 0.04 0.07\\
NGC 362 &62&-1.05 0.04 0.13&          -1.09 0.04 0.16&          0.15 0.06 0.09&          0.26 0.07 0.08&          0.29 0.06 0.09&          0.26 0.06 0.09\\
M 2 &78&-1.47 0.06 0.09&          -1.54 0.06 0.12&          0.17 0.11 0.09&          0.34 0.13 0.08&          0.23 0.07 0.10&          0.25 0.06 0.09\\
NGC 6752 &110&-1.48 0.06 0.09&          -1.56 0.07 0.12&          0.20 0.09 0.09&          0.35 0.11 0.09&          0.17 0.07 0.09&          0.23 0.07 0.10\\
NGC 1904 &44&-1.51 0.05 0.09&          -1.60 0.07 0.12&          0.16 0.09 0.09&          0.31 0.11 0.09&          0.21 0.08 0.10&          0.24 0.09 0.10\\
NGC 4833 &33&-1.88 0.06 0.08&          -2.08 0.08 0.11&          0.18 0.17 0.08&          0.36 0.20 0.10&          0.22 0.06 0.08&          0.24 0.07 0.10\\
NGC 4372 &45&-2.07 0.06 0.09&          -2.34 0.08 0.13&          0.31 0.07 0.09&          0.51 0.09 0.09&          0.20 0.06 0.08&          0.22 0.07 0.10\\
M 15 &46&-2.28 0.06 0.08&          -2.58 0.07 0.10&          0.22 0.19 0.11&          0.36 0.23 0.09&          0.21 0.05 0.09&          0.19 0.05 0.12\\
\hline
    \end{tabular}
    \label{tab:newnlte}
\end{table*}

\section{Discussion}
\label{discussion}
\subsection{Comparison with the literature}
\par
In what follows, we discuss our results for the Galactic clusters in the context of their chemical properties.  Many literature abundances are given in ``standard'' format: mean $\pm$ intra-cluster spread, computed as a simple standard deviation using all measurement in the cluster. In some cases, when not given in the same format, we recompute the mean and the standard deviations using the values of individual stars in the literature. Our own results are presented in the same format with mean from ML analysis and the {\em observed} intracluster spread (not the ML estimated internal dispersion) given in Table~\ref{tab:newnlte}.

We start with two open clusters and then continue with globular clusters, in order from the most metal-rich to the most metal-poor one.
\subsubsection{NGC~3532}
NGC 3532 is a young nearby metal-rich cluster at a heliocentric distance of $d_\odot \sim 0.5$ kpc \citep{Clem2011, Fritzewski2019}. The cluster has been extensively surveyed for  white dwarfs \citep{Dobbie2009,Dobbie2012} which has allowed accurate estimate of the cluster age of $\sim 300$ Myr from the white dwarf cooling sequence.%

On the basis of $12$ main-sequence stars, we find the metallicity $\feh_{\rm NLTE}=-0.10\ \pm\ 0.02$ dex and $\feh_{\rm LTE} =-0.09\ \pm\ 0.03$~dex. This estimate is consistent, within the uncertainties, with estimates based on the analysis of high-resolution spectra by \citet{santos2012}, \citet{Conrad2014c}, and \citet{Netopil2017c}. \citet{Fritzewski2019} reported the metallicity of $\feh $ of $-0.07\ \pm\ 0.10$ dex using lower-resolution near-IR spectra. 

Our NLTE abundance ratios suggest that the cluster is moderately $\alpha$-poor, with $\mgfe_{\rm NLTE}$ of $-0.09\ \pm\ 0.01$~dex,  although the [Ti/Fe] ratio is solar $\tife_{\rm NLTE} =0.01\ \pm\ 0.03$~dex. The [Mn/Fe] ratio is sub-solar, $\mnfe_{\rm NLTE}=-0.16\ \pm\ 0.03$~dex.
\subsubsection{NGC 2243}
\label{ngc2243}
NGC 2243 is an old Galactic open cluster located below the Galactic plane, at z $=-1.1$ kpc, and at a Galactocentric distance of $10.7$ kpc \citep{jacobson2011}. The age of the cluster was determined by several methods including spectroscopy, CMD isochrone fitting \citep{Anthony-Twarog2005}, using model age-luminosity and age-radius relations for eclipsing binaries \citep{Kaluzny2006}, bracketing $4 \pm 1$ Gyr.

The cluster has been subject to a very detailed chemical abundance analysis (for example the review by \citealt{Heiter2014}). \citet{Gratton1982} and \citet{Gratton1994} derived a spectroscopic metallicity of $\feh = -0.42\ \pm\ 0.05$, as well as detailed chemical abundances of the elements from C to Eu for a few RGB stars in the cluster. Their estimates were confirmed by \citet{Friel2002} and \citet{jacobson2011}, who  derived Fe, Ni, Ca, Si, Ti, Cr, Al, Na, and Mg abundances in a small sample of RGB stars. According to the latter study, this is one of the most metal-poor clusters at its location at an R$_{\rm GC} \sim 11$ kpc. This cluster has also been observed within the OCCAM APOGEE survey \citep{cuhna2016}. Their estimates of abundances in NGC 2243 are somewhat different from \citet{jacobson2011}, with Mg being $-0.14$ dex lower and more subtle differences for the other elements. In contrast to \citet{jacobson2011}, \citet{cuhna2016} also find a very large spread of metallicities in the cluster members, ranging from $-0.4$ to $+0.3$ dex.

\citet{Franccois2013} reported detailed abundances for the main-sequence and subgiant stars in the cluster. Their $\feh$ of $-0.54\ \pm\ 0.10$ dex is consistent with our NLTE estimate of $\feh_{\rm NLTE}=-0.52\ \pm\ 0.06$~dex. Our estimate of $\tife_{\rm NLTE} =0.02\ \pm\ 0.08$~dex is also in agreement with the value obtained by \citet{Franccois2013}, $\tife = 0.20\ \pm\ 0.22$ dex, within the combined uncertainties of both measurements. In fact, our lower estimate of $\tife$ corroborates the scaled-solar estimates of other $\alpha$-elements reported by \citet{Franccois2013}, [Ca/Fe] $=0.00\ \pm\ 0.14$ dex and [Si/Fe] $=0.12\ \pm\ 0.20$ dex. %

\subsubsection{NGC 5927}

NGC 5927 is a metal-rich globular cluster located close to the Galactic plane, at an altitude $z \sim 0.6$ kpc \citep{Casetti-Dinescu2007}. With an age of $12$ Gyr \citep{Dotter2010} and metallicity of \feh $\sim-0.5$ dex \citep{muraguzman2018}, the cluster is among the oldest metal-rich clusters known in the Galaxy. High-resolution spectroscopy of the cluster revealed the presence of multiple populations, especially prominent in the anti-correlation between Na and O \citep{Pancino2017,muraguzman2018}. The latter study also pointed out a similarity in the chemical properties of NGC 5927 and NGC 6440, a metal-rich GC in the Galactic bulge that could potentially hint at the common origin of the both systems.%

Our NLTE estimate $\feh_{\rm NLTE}=-0.48\ \pm\ 0.05$~dex is in very good agreement with earlier spectroscopic studies \citep[][\feh $=-0.47\ \pm\ 0.02$ dex]{muraguzman2018}. However, the abundance ratios are somewhat different. In particular, we find both Mg and Ti to be higher, $\mgfe_{\rm NLTE}=0.39\ \pm\ 0.04$~dex and $\tife_{\rm NLTE}=0.29\ \pm\ 0.06$~dex, compared to the results of the latter study. For Ti, our higher estimate is likely the consequence of NLTE over-ionisation, as the LTE abundance is $\tife_{\rm LTE}=0.23\ \pm\ 0.05$~dex, which is consistent with the estimate of $\tife=0.32\ \pm\ 0.05$~dex from \citet{muraguzman2018}. In contrast, the difference in Mg abundance is not related to NLTE. Our LTE Mg abundance is $\mgfe_{\rm LTE}=0.41\ \pm\ 0.05$~dex, which is much higher than that of \citet{muraguzman2018}, $\mgfe = 0.27\ \pm\ 0.02$ dex. It is possible that the differences stem from the differences in atomic data and/or model atmospheres. \citet{muraguzman2018} employ the MOOG code, Kurucz model atmospheres, and linelists from \citet{Villanova2011} and references therein. Our linelists have been extensively updated over the past years, and in particular for Mg lines, we used the data from \citet{PehlivanRhodin2017}. We were unable to find the atomic data in \citet{Villanova2011} and hence cannot provide a detailed analysis of the consistency of the models.
Our average [Mn/Fe] abundance ratio in NGC 5927 is sub-solar $\mnfe_{\rm NLTE}=-0.20\ \pm\ 0.03$~dex, $\mnfe_{\rm LTE}=-0.34\ \pm\ 0.03$~dex. This estimate is much lower compared to $\mnfe=-0.09\ \pm\ 0.08$ dex derived by \citet{muraguzman2018}, but it is mostly due to the difference of $-0.16$ in the adopted solar abundance (logA(Mn)$_\odot=5.37$~dex and logA(Mn)$_\odot=5.53$~dex, respectively).
\subsubsection{NGC~104 (47 Tuc)}

NGC~104 (47 Tuc) is among the brightest and most well-studied clusters in the Milky Way \citep[e.g.][] {Anderson2009,Campos2018u,milone2012,Lapenna2014,Cordero2014,thygesen2014,vCerniauskas2017}. The recent estimate of the distance to the cluster is $d_\odot = 4.45$ kpc \citep{Chen2018}, which was obtained on the basis of Gaia DR2 parallaxes. The reddening towards the system is very low $E(B-V)=0.03\ \pm\ 0.10$ mag allowing an accurate  estimate of the cluster age of $\sim 12.5$ Gyr \citep{brogaard2017}. Chemical abundance patterns, in the form of Na-O anti-correlations, enrichment in He and N, and depletion of C, indicate complex chemical evolution in the cluster \citep{Cordero2014,Kuvcinskas2014,Marino2016j}.

Our NLTE estimate of the cluster metallicity, $\feh_{\rm NLTE}=-0.74\ \pm\ 0.03$~dex, is in very good agreement with previous estimates \citep{Koch2008f, Cordero2014, Dobrovolskas2014,thygesen2014}. The latter study reports $\feh=-0.78\ \pm\ 0.07$ dex obtained by 1D LTE modelling of Fe lines. The authors also test the effect of NLTE, finding the effects to be of the order $+0.02$ dex on the Fe abundances. Indeed, this is fully confirmed by our LTE metallicities, which are $0.01$ dex lower compared to our NLTE results. For Mg, \citet{thygesen2014} report [Mg/Fe] $=0.44\ \pm\ 0.05$ dex in LTE, which is in excellent agreement with our LTE value,  $\mgfe_{\rm LTE}=0.42\ \pm\ 0.04$~dex, and is only slightly higher than our NLTE result  $\mgfe_{\rm NLTE}=0.38\ \pm\ 0.05$~dex. Also the Ti abundances are consistent with \citet{thygesen2014}. We obtain  $\tife_{\rm NLTE}=0.30\ \pm\ 0.07$~dex and $\tife_{\rm LTE}=0.26\ \pm\ 0.07$~dex, which agrees within the uncertainties with the measured value of [Ti/Fe]=$0.28\ \pm\ 0.08$ dex from \citet{thygesen2014}.
%
%
%
\subsubsection{NGC 1851}
NGC 1851 is a moderately metal-poor globular cluster located at an $R_{\rm GC}$ of 17 kpc from the Galactic centre and $\sim 7$ kpc below the disk plane \citep[][2010 edition]{harris}. \citet{wagnerkaiser2017} find a cluster age of $11.5$ Gyr. Some have argued for an evolutionary connection between NGC 1851 and several other clusters (NGC 1904, NGC 2808, and NGC 2298) on the basis of their spatial proximity \citep{Bellazzini2001}, as we confirm by our abundances below. An idea has been put forward that all four clusters are associated with the disrupted Canis Major dwarf galaxy \citep{martin2004}. Others suggest that NGC 1851 is possibly the nucleus of a disrupted dwarf galaxy \citep{Bekki2012,Kuzma2018} or could have formed as a result of the merger of two globular clusters \citep{carretta2011}. The cluster hosts multiple stellar populations, seen in photometric data on the main sequence, subgiant branch, and RGB \citep{Milone2008,turri2015,Cummings2017}. Also the spectroscopic analysis of C and N suggests the presence of several populations \citep{Yong2008,yong2015,Simpson2017}.

Our metallicities for NGC 1851 are slightly higher compared to previous studies. \citet{Gratton2012a} find a range of metallicities in the cluster from $\feh=-1.23\ \pm\ 0.06$ dex (subgiant branch) to $\feh=-1.14\ \pm\ 0.06$ dex (RGB). Our analysis yields $\feh_{\rm NLTE}=-1.11\ \pm\ 0.04$~dex and $\feh_{\rm LTE}=-1.15\ \pm\ 0.04$~dex, whereas \citet{yong2015} report $\feh = -1.28\ \pm\ 0.05$ and \citet{Marino2014} obtain $\feh = -1.33\ \pm\ 0.09$ dex.

For Mg, we find  $\mgfe_{\rm NLTE} =0.22\ \pm\ 0.08$~dex, which is lower than the value reported by \citet{Marino2014} $\mgfe = 0.44\ \pm\ 0.16$ dex. However, this difference can be almost entirely explained by NLTE. Indeed our LTE estimates of $\mgfe$ are much higher, $\mgfe_{\rm LTE} =0.36\ \pm\ 0.05$~dex, and are also in agreement with the LTE estimates by \citet{carretta2011}, $\mgfe = 0.35\ \pm\ 0.03$ dex. For Ti, we find the opposite offset, in the sense that our NLTE values, $\tife_{\rm NLTE} =0.28\ \pm\ 0.06$~dex, are higher compared to the LTE results by \citet{carretta2011} $\tife=0.17\ \pm\ 0.05$ dex. This can be explained by NLTE, as our LTE abundances of Ti are slightly lower, $\tife_{\rm LTE} =0.24\ \pm\ 0.06$~dex, consistent with the latter study within the combined uncertainties of the both LTE measurements.

It is interesting, in the context of the common formation scenario of NGC 1851 and NGC 2808, as proposed by \citet{martin2004}, that our chemical abundances in the two clusters are very similar. In fact, given the uncertainties of our measurements, both clusters are consistent with being formed from the same material, and having the same progenitor system.
\subsubsection{NGC 2808}
NGC 2808, a moderately metal-poor old cluster, is among the most massive and complex systems in the Milky Way galaxy \citep{Simioni2016}, with the mass of $7.42 \times 10^5$ M$_\odot$ \citep{Baumgardt2018} and multiple populations \citep{Piotto2007,Milone2015}. NGC 2808 was among the first clusters, for which a prominent Na-O anti-correlation was reported \citep{Carretta2006}, along with a He spread \citep{DAntona2005} and a Mg-Al anti-correlation \citep{Carretta2006b}.

Our LTE metallicity, $\feh_{\rm LTE} =-1.03\ \pm\ 0.05$~dex, is slightly higher compared to the recent literature values. \citet{carretta2015} report $\feh=-1.13\ \pm\ 0.03$~dex using the Fe I lines and $\feh=-1.14\ \pm\ 0.03$~dex using the Fe II lines. They also find a large spread in [Mg/Fe] abundance ratios, which is corroborated by our results. In particular, we find that the individual LTE abundance ratios of [Mg/Fe] range from $0.08$ to $0.45$ dex, and the average value and its dispersion, $\mgfe_{\rm LTE}=0.22\ \pm\ 0.15$~dex, is consistent with  $\mgfe=0.26\ \pm\ 0.16$~dex obtained by \citet{carretta2015}. For Ti, our estimate $\tife_{\rm LTE}=0.29\ \pm\ 0.04$~dex is slightly higher compared to   $\tife=0.21\ \pm\ 0.04$ dex derived by \citet{carretta2015}. Our NLTE measurements are: 
$\feh_{\rm NLTE} =-1.01\ \pm\ 0.05$~dex, $\mgfe_{\rm NLTE} =0.11\ \pm\ 0.14$~dex, and $\tife_{\rm NLTE} =0.33\ \pm\ 0.04$~dex.
\subsubsection{NGC 362}

The globular cluster NGC 362 has been extensively studied in the literature since the early work by \citet{Menzies1967}. A recent analysis of Gaia~DR2 astrometric data by \citet{Chen2018} places it at a heliocentric distance of $8.54$~kpc, relatively close to the Galactic disk plane. Photometric studies of the cluster revealed multiple sequences on the HB \citep{Dotter2010,Gratton2010a,Piotto2012}. Spectroscopic follow-up confirmed its unique nature, with discrete groups of Na/O ratios \citep{carretta2013}, a bimodal distribution of CN \citep{Smith2009,Lim2016}, a very large spread of Al abundances.

Our NLTE metallicity for this cluster, $\feh_{\rm NLTE}=-1.05\ \pm\ 0.04$~dex, is somewhat higher compared to the results of the earlier studies. Our LTE estimate is lower, $\feh_{\rm LTE} =-1.09\ \pm\ 0.04$~dex and is consistent with the RR Lyr-based value from \citet{Szekely2007}. A very careful analysis of high-resolution spectra by \citet{Worley2010} yielded $\feh  = -1.20\ \pm\ 0.09$ (from the Fe II lines), which is consistent within the uncertainty with our LTE estimate. A somewhat lower value is reported by \citet{dorazi2015}. They find $\feh$ of $-1.26$ dex from the LTE analysis of RGB stars. The perhaps most extensive chemical study of the cluster, to date, is that by \citet{carretta2013} employing UVES and Giraffe spectra of 138 RGB stars. For the UVES sample, they find a mean LTE metallicity of $\feh =-1.17\ \pm\ 0.05$ dex from the Fe I lines and $\feh =-1.21\ \pm\ 0.08$ dex from Fe II lines that is in agreement with our LTE metallicity. Their abundance of [Ti/Fe] ($0.22\ \pm\ 0.04$ dex based on the UVES spectra) and [Mg/Fe] ($0.33\ \pm\ 0.04$ dex) are also in good agreement with our LTE estimates, $\tife_{\rm LTE} =0.26\ \pm\ 0.06$~dex and  $\mgfe_{\rm LTE} =0.26\ \pm\ 0.06$~dex. In contrast, our NLTE values are considerably different, $\tife_{\rm NLTE} =0.29\ \pm\ 0.06$~dex and $\mgfe_{\rm NLTE} =0.15\ \pm\ 0.06$~dex. To the best of our knowledge, this paper is the first study to provide estimates of NLTE abundances in this cluster.
\subsubsection{M2 (NGC 7089)}
M2 is an old cluster in the Galactic halo at a distance of $\sim 7$ kpc below the Galactic plane and at a heliocentric distance of 11.5 kpc \citep[][2010 edition]{harris}. The cluster was the first system, in which a CN distribution bimodality was detected \citep{Smith1990,Lardo2012,Lardo2013}.  \citet{Yong2014} argued for a trimodal metallicity distribution that has been, however, disputed by \citet{Lardo2016}, who found a bimodal distribution using Fe II lines. \citet{Milone2015} employed HST photometry to detect a very rich stellar environment, composed of three main populations distinguished by their metallicity and a spread in He abundance from the primordial mass fraction of $Y\sim 0.25$ to $Y\sim 0.31$. They also suggest that there are six sub-populations with unique light element abundance patterns, that could potentially hint at either an independent enrichment and star formation history of the individual components or at a unique merger formation history of the cluster. The imaging data by \citet{Kuzma2016} further strengthen the latter interpretation, by demonstrating a diffuse stellar envelope that could possibly indicate that the GC is a stripped dSph nucleus.

We find a modest metallicity spread in the cluster $\feh_{\rm NLTE}=-1.47\ \pm\ 0.06$~dex. Our LTE result $\feh_{\rm LTE}=-1.54\ \pm\ 0.06$~dex is in good agreement with the previous measurements, in particular with \citet{Lardo2016}, who derive $\feh =-1.50\ \pm\ 0.05$ dex for the metal-poor component, using Fe II lines.  \citet{Yong2014} report three groups with [Fe/H] ranging from $-1.66\ \pm\ 0.06$~dex to $-1.02\ \pm\ 0.06$ dex, as derived from the Fe II lines. It should be noted, however, that \citet{Lardo2016} suggest that the metal-rich component may not constitute more than 1 \% of the cluster population. As to abundance ratios, comparing our LTE estimates with \citet{Yong2014}, we find a good agreement in Mg with $\mgfe_{\rm LTE} =0.34\ \pm\ 0.13$~dex, that should be compared to their estimates of $0.38\ \pm\ 0.08$ dex. Yet, similar to the other clusters, our NLTE abundance of Mg is lower, $\mgfe_{\rm NLTE} =0.17\ \pm\ 0.11$~dex. We obtain $\tife_{\rm NLTE} = 0.23\ \pm\ 0.07$~dex in NLTE, and $\tife_{\rm LTE} =0.25\ \pm\ 0.06$~dex in LTE, which is lower than the estimates derived by \citet{Yong2014} $\tife=0.31\ \pm\ 0.12$ dex. We note, however, that their approach leads to a significant ionisation imbalance of Ti I - Ti II in the two groups, and it is not clear which of the estimates is more reliable. Our measurement of [Ti/Fe] is more consistent with their estimate based on the Ti II lines.
\subsubsection{NGC 6752}
NGC 6752 belongs to the benchmark globular clusters in our Galaxy. Its proximity, $d_\odot$ of $4.0$ kpc \citep[][2010 edition]{harris}, allows a detailed spectroscopic and photometric analysis of the cluster members. The cluster has been extensively observed with the VLT \citep[e.g.][]{Carretta2007,Gruyters2014,Lee2018} and with HST \citep[e.g.][]{Ross2013,Gruyters2017,Milone2019}. In particular, deep narrow-band photometric observations have been essential to probe the substructure of this system, with multiple stellar populations identified on the RGB and MS \citep{Dotter2015,Lee2018,Milone2019}. %

A detailed chemical analysis of the cluster members was presented in different studies. The analysis of high-resolution UVES spectra of 38 RGB stars in NGC 6752 by \citet{Yong2005} showed a prominent $\alpha$-enhancement at $ \mgfe=0.47\ \pm\ 0.06$ dex, and the iron abundances of $\feh=-1.56\ \pm\ 0.10$ dex. Both of these estimates are fully consistent with our LTE results of $\feh_{\rm LTE}=-1.56\ \pm\ 0.07$~dex and $\mgfe_{\rm LTE}=0.35\ \pm\ 0.11$~dex. Furthermore, their LTE estimate of Ti abundance, $\tife=0.14\ \pm\ 0.14$ dex, is consistent with our LTE value, $\tife_{\rm LTE}=0.23\ \pm\ 0.07$~dex. Our sample is larger than that of \citet{Yong2005} and comprises $110$ stars at the base of the RGB, which may account for minor differences between our and their results. On the other hand, our somewhat larger dispersion in abundance ratios is probably not an artefact, as large intra-cluster abundance spreads have also been reported by \citet{Yong2013} from the analysis of  high-resolution spectra of RGB stars. Our NLTE estimates are slightly different, but they follow the general trends identified for other metal-poor clusters. The NLTE metallicity and slightly higher, $\feh_{\rm NLTE}=-1.48\ \pm\ 0.06$~dex, whereas the NLTE $\mgfe$ ratio is correspondingly lower, $\mgfe_{\rm NLTE}=0.20\ \pm\ 0.09$~dex. 
\subsubsection{NGC 1904 (M79)}
NGC 1904 is a metal-poor globular cluster at $d_\odot =$ 12.9 kpc and 6.3 kpc below the Galactic plane \citep[][2010 edition]{harris}. \citet{kains2012} employed variable stars to determine an accurate distance to the cluster, 13.4$\pm$0.4 kpc. The age of the system is 14.1$\pm$2.1 Gyr \citep{li2018}. Similar to NGC 1851, the outskirts of NGC 1904 reveal prominent streams signifying its possible accretion origin \citep{Carballo-Bello2018,Shipp2018}. 

Our NLTE metallicity of the cluster is $\feh_{\rm NLTE}=-1.51\ \pm\ 0.05$~dex. This is consistent, modulo the LTE - NLTE difference of -0.07 dex, with the value reported by \citet{carretta2009}, $\feh=-1.58\ \pm\ 0.03$ dex. Also their LTE Mg abundance, $\mgfe=0.28\ \pm\ 0.06$ dex, is in good agreement with our LTE value of $\mgfe_{\rm LTE}=0.31\ \pm\ 0.11$~dex. Our NLTE estimate is $\mgfe_{\rm NLTE}=0.16\ \pm\ 0.09$~dex, which is lower than the LTE value. The cluster is also enriched in [Ti/Fe]. We find $\tife_{\rm NLTE}=0.21\ \pm\ 0.08$~dex and $\tife_{\rm LTE}=0.24\ \pm\ 0.09$~dex, and the latter is consistent with the LTE results obtained by \citet{Fabbian2005}, $\tife=0.31\ \pm\ 0.15$ dex.
\subsubsection{NGC 4833}
The cluster is arguably one of the oldest systems in the Milky Way, with the age of 13.5 Gyr \citep{wagnerkaiser2017}. Its location at $d_\odot = 6.6$ kpc, $\sim 1$ kpc away from the disk plane \citep[][2010 edition]{harris} and orbital eccentricity are consistent with the cluster being a part of the inner halo system \citet{Carretta2010b}. The cluster is thought to host multiple populations \citep{Carretta2014a}, based on chemical signatures.

A detailed spectroscopic analysis of the cluster has been performed by several groups. \citet{Carretta2014a} employed UVES and Giraffe spectra of 78 stars to determine the abundances of $20$ elements from Na to Nd. They obtained relatively small dispersions for the majority of elements, including Fe. In contrast, they also found very pronounced Na-O and Mg-Na anti-correlations and a large intra-cluster variation in the abundances of light elements. Specifically, the [Mg/Fe] abundance ratios in the cluster range from slightly sub-solar, [Mg/Fe] $\sim -0.05$ dex, to highly super-solar values, [Mg/Fe] $\gtrapprox 0.7$ dex.  Another high-resolution study of the cluster was presented by \citet{roediger2015}, who obtained high $\snr$ spectra with the MIKE spectrograph at the Magellan II telescope. Their estimates of elemental abundances are somewhat different from \citet{Carretta2014a}. In particular, they report $\feh=-2.25\ \pm\ 0.02$ dex from the neutral Fe lines, and $\feh=-2.19\ \pm\ 0.01$ dex from the ionised Fe lines, attributing the differences with respect to \citet{Carretta2014a} to the technical aspects of the analysis, such as the the linelist and the solar reference abundances. In terms of abundance inhomogeneities and correlations, their study is consistent with \citet{Carretta2014a}, with pronounced star-to-star variations in the light elements and signatures of bimodality in Na, Al, and Mg.  

Our LTE estimates of metallicity and abundance ratios are consistent with the literature estimates. In particular, we find $\feh_{\rm LTE}=-2.08\ \pm\ 0.08$~dex and $\mgfe_{\rm LTE}=0.36\ \pm\ 0.20$~dex, which can be compared to $\feh=-2.04\ \pm\ 0.02$ dex and $\mgfe = 0.36\ \pm\ 0.15$ dex derived by \citet{Carretta2014a} from the Giraffe spectra. We also confirm that there is negligible internal dispersion in Ti abundances, with $\tife_{\rm LTE}$ of $0.24\ \pm\ 0.07$~dex, consistent with \citet{Carretta2014a} estimate of $\tife = 0.17\ \pm\ 0.02$ dex. On the other hand, our NLTE abundances are considerably different. For Fe, we infer $\feh_{\rm NLTE}=-1.88\ \pm\ 0.06$~dex, which is higher compared to $\feh_{\rm LTE}=-2.08\ \pm\ 0.08$~dex. Also, the [Mg/Fe] ratios are much lower, $\mgfe_{\rm NLTE}=0.18\ \pm\ 0.17$~dex, with  the abundances in the individual stars ranging from $-0.03$ to $0.70$ dex. The NLTE Ti abundances are only slightly higher compared to the LTE estimates, $\tife_{\rm NLTE} =0.22\ \pm\ 0.06$~dex. %
\subsubsection{NGC 4372}
Remarkable for its strong chemical peculiarities, NGC 4372 is nonetheless a rather typical GC system. It is metal-poor and nearby cluster, with an age of 12.5 Gyr \citep{kruijssen2018}, at a distance of $5.8$ kpc and $1.0$ kpc below the Galactic plane \citep[][2010 edition]{harris}.  The cluster reveals a signficant dispersion in Na, Mg, Al, and O, a Na-O anti-correlation, and, possibly, an Al-Mg anti-correlation \citep{sanroman2015}.%

Our average NLTE metallicity of stars in NGC 4372 is $-2.07\ \pm\ 0.06$~dex. Our LTE metallicity is much lower, $\feh_{\rm LTE} =-2.33\ \pm\ 0.08$~dex, following the general trend for all metal-poor clusters seen in Fig.\ref{fig:4}. Comparing the latter estimate with the literature, we find a satisfactory agreement with a comprehensive study by \citet{sanroman2015}, which is also based on the spectra acquired within the Gaia-ESO survey. Their estimate of $\feh$ is $-2.23\ \pm\ 0.10$ dex\footnote{Note that this value depends on whether large outliers are included or not.}, consistent with our results within the combined uncertainties of the both estimates. Also the value from \citet{carretta2009}, $\feh = -2.19\ \pm\ 0.08$ dex, is somewhat higher than our LTE metallicity. The detailed abundance ratios of our study are also in agreement with those measured by \citet{sanroman2015}. We obtain $\mgfe_{\rm LTE} =0.51\ \pm\ 0.09$~dex and $\tife_{\rm LTE} =0.22\ \pm\ 0.07$~dex in LTE, whereas \citet{sanroman2015} derive $\mgfe = 0.44\ \pm\ 0.07$ dex and $\tife = 0.31\ \pm\ 0.03$ dex.  Our NLTE estimates are $\mgfe_{\rm NLTE} =0.31\ \pm\ 0.07$~dex and $\tife_{\rm NLTE} =0.20\ \pm\ 0.06$~dex.
\subsubsection{M15 (NGC 7078)}
Similar to NGC 1904 and NGC 4833, M15 represents one of the oldest and metal-poor systems in the Galactic halo at a distance $d_\odot = 10.4$ kpc and $4.8$ kpc below the Galactic plane \citep[][2010 edition]{harris}. Several studies report multiple stellar populations in the cluster \citep{Larsen2015,Nardiello2018,Bonatto2019}.  

M15 has the lowest metallicity in our sample and shows the largest NLTE effects: $\feh_{\rm NLTE}=-2.28\ \pm\ 0.06$~dex, but $\feh_{\rm LTE}=-2.58\ \pm\ 0.07$~dex. Our LTE estimate compares favourably well with \citet{sobeck2011}, who derived $\feh = -2.62\ \pm\ 0.08$ dex\footnote{We recompute value using the mean of all measurements from nine RGB and RHB stars.} from the analysis of high-resolution spectra of several RGB and RHB stars in the cluster collected with the HIRES spectrograph at the Keck telescope. \citet{Worley2013} report $\feh$ in the range from $-2.4$ to $-2.3$ dex with an uncertainty of $0.1$ dex, which is closer to the estimate of $\feh=-2.37$ dex derived by \citet{letarte2006} and $\feh=-2.32$ dex by \citet{carretta2009}. Our average LTE abundances of Mg is $\mgfe_{\rm LTE}=0.36\ \pm\ 0.23$~dex,  with the star-to-star variation in the range from $-0.26$ to $0.66$ dex. This is consistent with \citet{carretta2009}, within the uncertainties, and also with the abundances derived by \citet{sobeck2011}, who measured [Mg/Fe] ratios from $-0.01$ to $0.6$~dex. In contrast, the cluster stars exhibit very tight [Ti/Fe] ratios with the mean of $\tife_{\rm LTE}=0.19\ \pm\ 0.05$ dex. Our NLTE results for Mg are much lower than the LTE ones, $\mgfe_{\rm NLTE}=0.22\ \pm\ 0.19$~dex, whereas the NLTE Ti abundances are nearly consistent with LTE, $\tife_{\rm NLTE}=0.21\ \pm\ 0.05$~dex.
\begin{figure}
    \centering
    \includegraphics[width=\columnwidth]{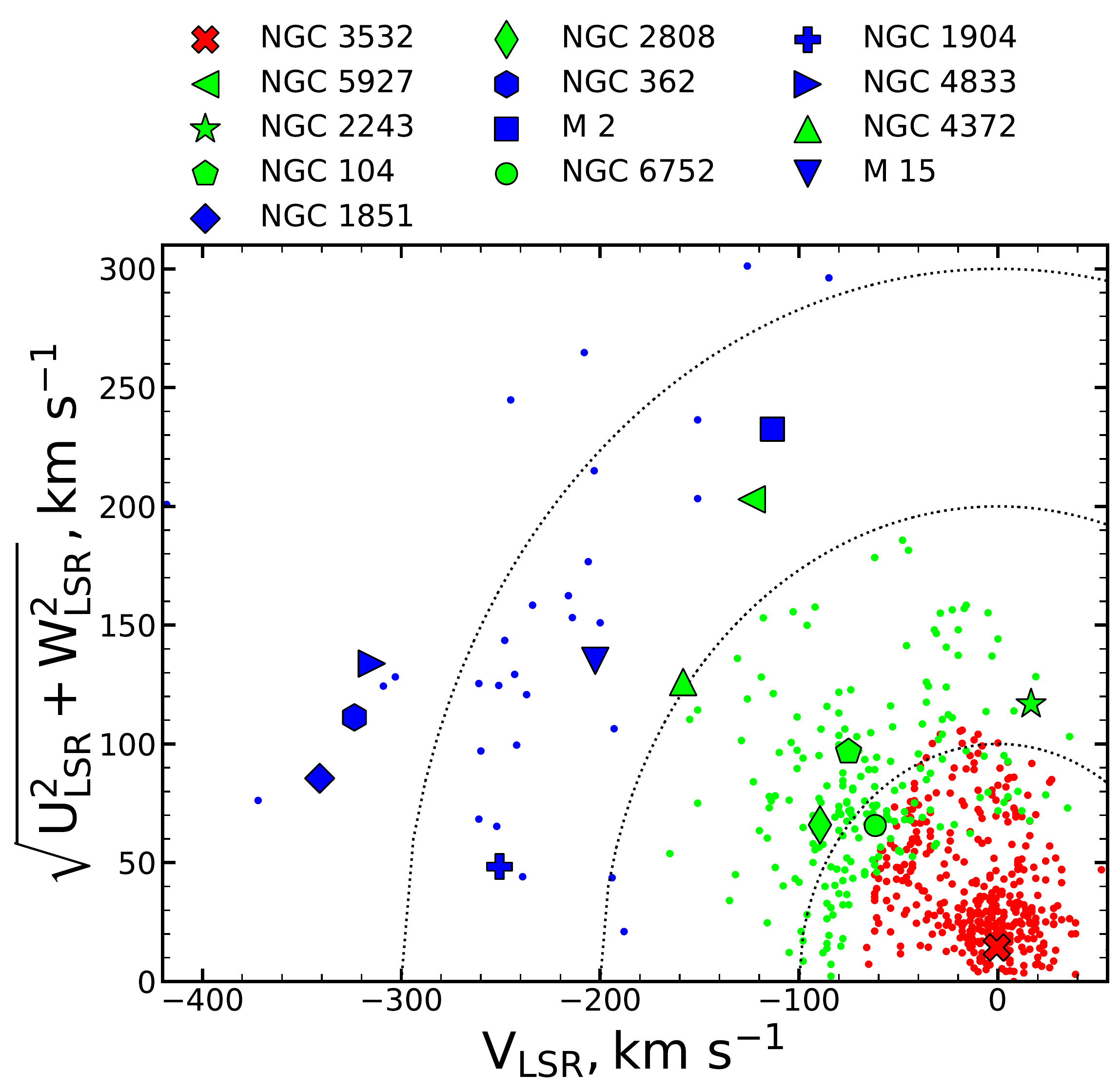}
    \caption{Toomre diagram for clusters and \protect\citet{Bensby2014} field stars. The thin disk population is shown in red colour, the thick disk population in green colour, and the halo population in blue colour. The isolines for total velocity $V_{\rm tot}=\sqrt{U_{\rm LSR}^2+V_{\rm LSR}^2+W_{\rm LSR}^2}=100,\ 200,\ 300~\kms$ are shown as dotted lines. For the details of population assignment see Appendix~\ref{distances}.}
    \label{fig:fig8}
\end{figure}
\begin{figure*}
\includegraphics[width=\textwidth]{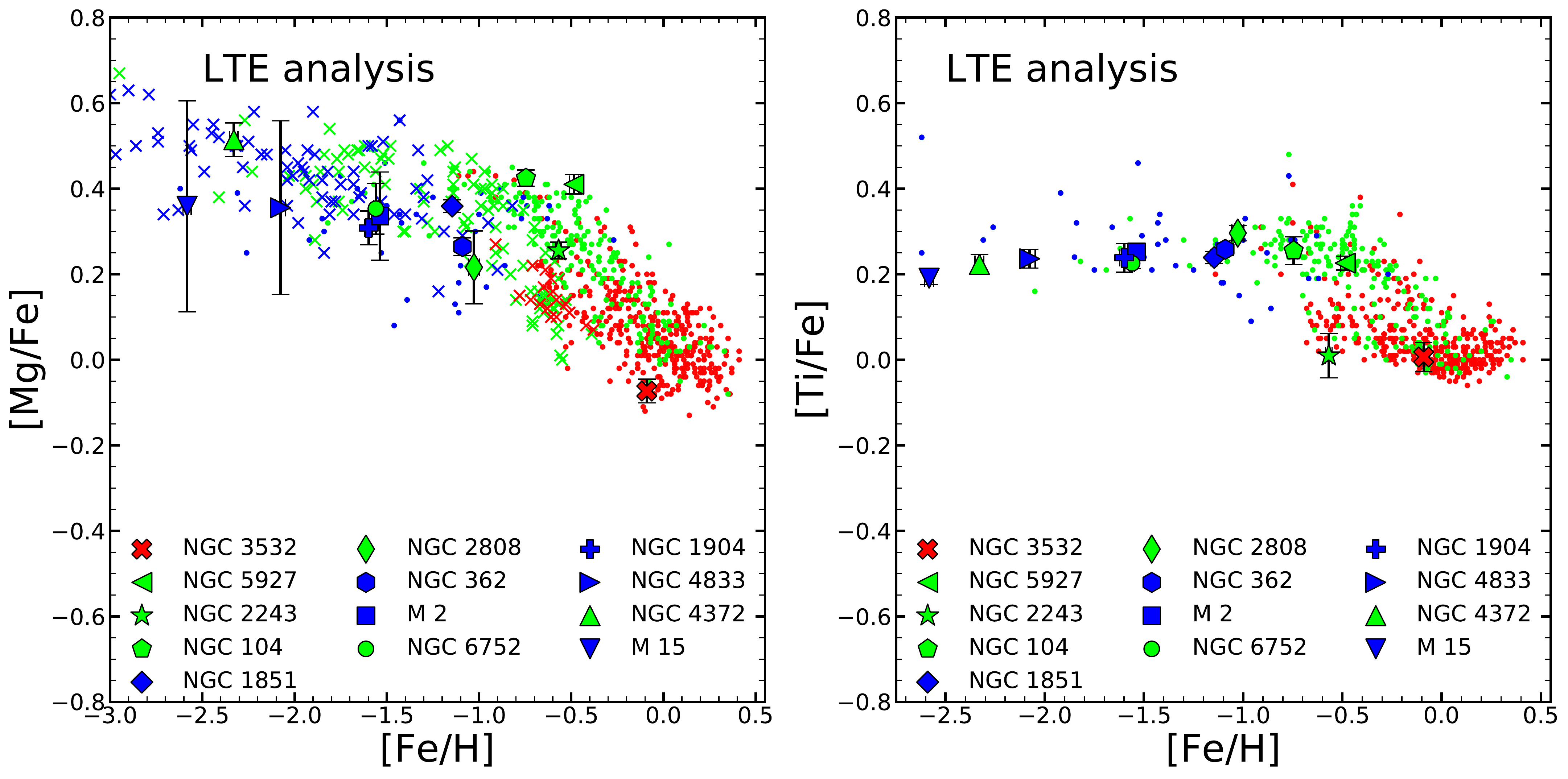}
\caption{Mean LTE metallicities and $\mgfe$ and $\tife$ abundance ratios for all clusters and for Milky Way field stars from \protect\citet{Bensby2014}(NLTE $\feh$, LTE $\mgfe$ and LTE $\tife$ -- small dots) and \protect\citet{bergemann2017b}(1D LTE results -- small crosses). Error bars represent the $1 \sigma$ intra-cluster abundance variations. Colours are the same as in Fig.~\protect\ref{fig:fig8}.}
\label{fig:fig9} 
\end{figure*}

\begin{figure*}
\includegraphics[width=\textwidth]{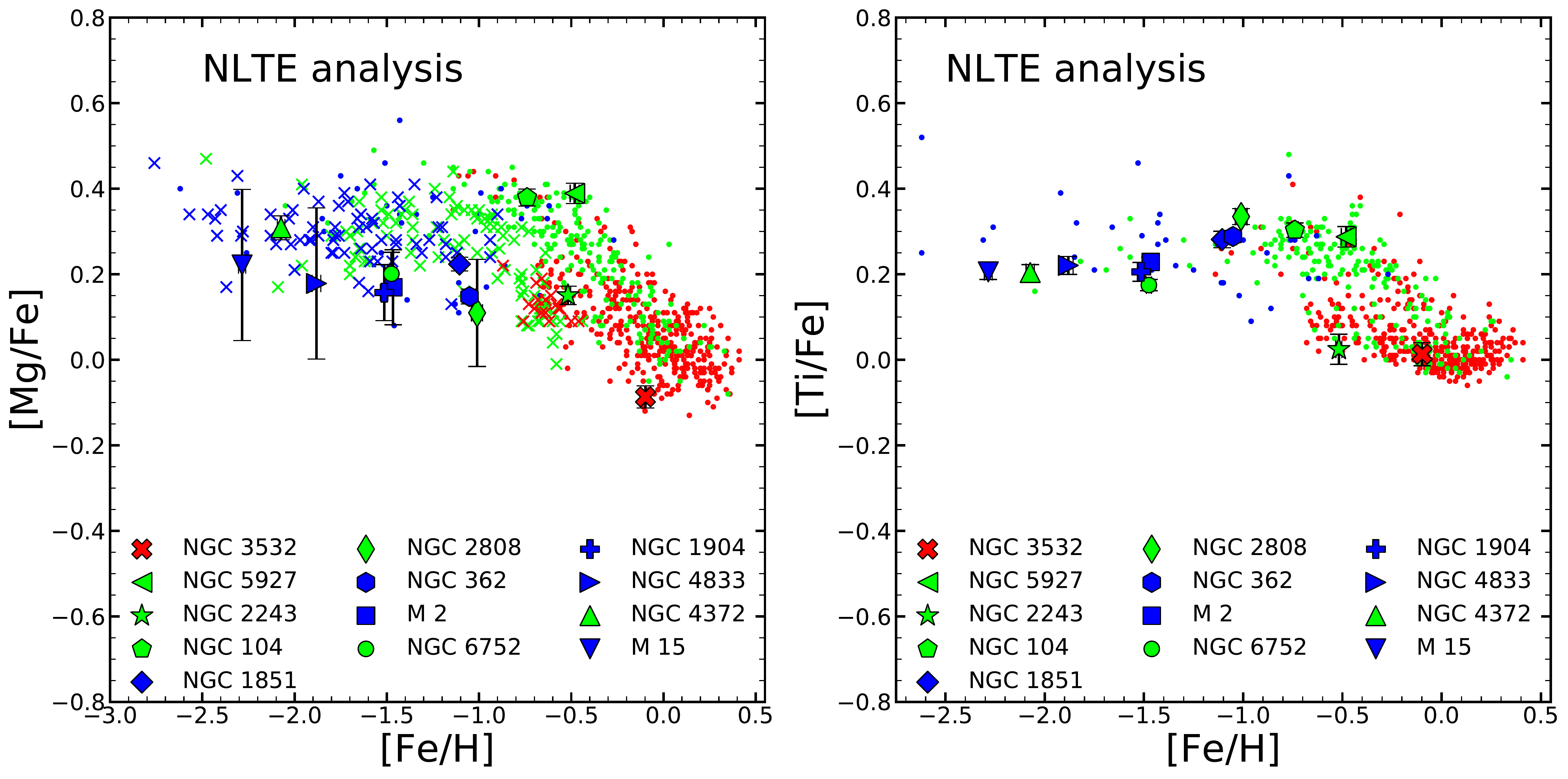}
\caption{Mean NLTE metallicities and $\mgfe$ and $\tife$ abundance ratios for all clusters and for Milky Way field stars from \protect\citet{Bensby2014}(NLTE $\feh$, LTE $\mgfe$ and LTE $\tife$ -- small dots) and \protect\citet{bergemann2017b}(1D NLTE results -- small crosses). Error bars represent the $1 \sigma$ intra-cluster abundance variations. Colours are the same as in Fig.~\protect\ref{fig:fig8}.}
\label{fig:fig10} 
\end{figure*}
\subsection{Comparison with Milky Way field stars}
It is useful to combine our chemical characterisation of the clusters with their kinematics, in order to compare our results with Galactic field stars. We employ the kinematic selection criteria from \citet{Bensby2014} to assign Galactic population membership to the clusters (see Appendix~\ref{distances}). The Toomre diagram for the clusters and field stars is shown in Fig.~\ref{fig:fig8}.

In Fig. \ref{fig:fig9} and \ref{fig:fig10}, we overlay our LTE and NLTE abundance ratios in the clusters with the literature measurements in the Galactic field stars. The field sample is taken from \citet{Bensby2014} and \citet{bergemann2017b}. The former dataset represents populations in the solar neighbourhood and has a large coverage in metallicity, $-2.7 \lesssim \feh \lesssim 0.5$. The Fe abundances were derived in NLTE, while Mg and Ti were derived in LTE analysis. The dataset \citet{bergemann2017b} lacks a thin disk component, $\feh > -0.5$, but contains a significant fraction of the thick disk and halo stars. The study provides LTE and NLTE estimates of $\feh$ and $\mgfe$ derived using 1D and <3D> atmospheric models. For consistency with our 1D analysis, we use their 1D LTE and 1D NLTE results.

There are several important results, which stand out by comparing our LTE and NLTE measurements in clusters against Galactic field stars. Firstly, our LTE abundances in GCs trace the Galactic field population remarkably well, at least as long as LTE field distributions are employed for the comparison. This supports the conclusions drawn by \citet{Pritzl2005}. NGC 3532 and NGC 2243, the two metal-rich clusters with disk-like kinematic properties, occupy the chemical locus of the thin disk. The metal-poor globular clusters trace the thick disk and the halo.
Despite a difference of  two orders of magnitude in metallicity, all metal-poor GCs follow very tight trends of the average [Mg/Fe] and [Ti/Fe] with [Fe/H]. In particular, all of them occupy the locus situated at [Ti/Fe] $\approx 0.25$ dex with small dispersion. On the other hand, the intra-cluster dispersions of [Mg/Fe] increase substantially. This is not unexpected and has been extensively discussed in the literature \citep{Gratton2004,Carretta2014a,Carretta2014b}. The large variation of Mg abundances is usually attributed to the nuclear processing associated with high temperature hydrogen burning and multiple star formation episodes. In such a scenario first generation massive stars evolve fast, converting their Mg into Al. Second generation stars, formed from the material of first generation stars, are depleted in Mg and enriched in Al. The absence of any noticeable dispersion in [Ti/Fe] in all GCs corroborates this interpretation.

Notwithstanding the good agreement of our LTE results with earlier LTE studies, we find important differences between LTE and NLTE results (Fig.~\ref{fig:fig10}), which impact the astrophysical interpretation of the results. When comparing our NLTE abundances for globular clusters with the NLTE abundances of field stars, only two metal-rich clusters with the thick disk kinematics (NGC 104 and NGC 5927) and the metal-poor cluster NGC~4372 appear to be consistent with the field stars. All other metal-poor clusters are systematically depleted in [Mg/Fe] relative to the metal-poor disk and the halo. This may imply that the metal-poor clusters were not formed \textit{in-situ}, but were accreted from disrupted dwarf satellite galaxies. %
\section{Conclusions}
\label{Conclusions}

In this work, we employ non-LTE radiative transfer models and the Payne code to determine chemical abundances for 13 stellar clusters in the Milky Way. The observed spectra are taken from the public 3$^{rd}$ data release of the Gaia-ESO survey, and we focus on the $R\sim 19\,800$ spectra taken with the Giraffe instrument. The NLTE synthetic spectra are computed using the model atoms presented in earlier works \citep{bergemann2008,Bergemann2011,Bergemann2012c,Bergemann2017a}. \textit{The Payne} code is used to interpolate in the grids of synthetic spectra to maximise the efficiency of the analysis, where we simultaneously fit for all spectral parameters, exploring more information from the full spectrum. The spectral grids are computed at random nodes in stellar parameter space and a $\chi^2$ minimisation is employed to find the best-fit stellar parameters and chemical  abundances by comparing the models with the observations.

We validate our method and the models on the Gaia-ESO benchmark stars, for which stellar parameters are well constrained by parallaxes, asteroseismology, and interferometric angular diameter measurements. The calibration sample includes $19$ main-sequence dwarfs, subgiants, and red giants in the $\feh$ range from $-2.5$ to $0.3$ dex with spectra taken at different exposure times spanning the $\snr$ range of 100 to 2800 \AA$^{-1}$. We find a very good agreement between our NLTE spectroscopic results and the independently determined stellar parameters. The residuals are within $-29 \pm 88$ K in $\teff$, $0.09\ \pm\ 0.16$ dex in $\logg$, and $0.02\ \pm\ 0.09$ dex in $\feh$. The analysis of repeat observations of the same stars indicates the absence of a systematic bias or correlation of the abundance error with the quality the spectra within the full range of $\snr$ probed in this work.

We compute stellar parameters and abundances for $742$ stars in two open clusters and $11$ globular clusters in the Milky Way galaxy. The results are provided in Table~\ref{tab:catalog} and are archived electronically on CDS. The typical $\snr$ of the spectra is 200~\AA$^{-1}$.  We find that spectroscopic estimates of stellar parameters ($\teff$, $\log g$, and $\feh$) agree with evolutionary expectations, based on isochrones. However, different isochrones are needed to match the LTE and NLTE data. At low metallicity, the difference between LTE and NLTE parameters is significant, confirming earlier studies \citep[i.e. ][]{Bergemann2012c,lind2012,ruchti2013}. The systematic error of LTE increases in proportionality with decreasing metallicity, and amounts to $300$ K in $\teff$, $0.6$ dex in $\log g$, and $0.3$ dex in $\feh$ for the RGB stars with $\feh_{\rm NLTE} = -2.3$.
The $\mgfe$ abundance ratios are typically lower in NLTE compared to LTE. Our abundances show no significant trends with stellar parameters, supporting their relative accuracy.

Our results for the Galactic open and globular clusters can be summarised as follows:

\begin{itemize}

\item NGC 3532, a young metal-rich open cluster, is consistent in its chemical abundance pattern and its kinematics with the Galactic thin disk. The cluster is slightly depleted in Mg compared to the solar neighbourhood, although the difference is generally within the uncertainties of the abundance measurements. 

\item NGC 2243, a relatively old open cluster lies on the metal-poor end of the thin disk track, and shows a noticeable dispersion in $\feh$, $\mgfe$, and $\tife$ ratios contrasting with the tight chemical patterns in the field stars. This is the only cluster in our sample that is represented by main-sequence and TO stars, and this spread likely has an astrophysical origin. In particular, the pronounced dip in $\feh$ at the TO signifies the action of atomic diffusion consistent with depletion predicted by detailed stellar evolution models.

\item Two metal-rich clusters with thick disk like kinematics NGC~104 and NGC~5927 are also very similar to the thick disk in their abundance ratios of [Mg/Fe] and [Ti/Fe]. They show small dispersions in all elements $\lesssim 0.06$~dex, which are much smaller then the typical systematic uncertainties of our measurements, and are consistent with being chemically homogeneous populations.

\item The metal-poor clusters NGC~2808 and NGC~6752, despite being kinematically similar to the thick disk, appear to be depleted in [Mg/Fe] compared to the field stars, based on NLTE analysis. On the other hand, their [Ti/Fe] ratios are representative of the halo clusters.

\item NLTE analysis suggests that the majority of metal-poor clusters with [Fe/H] $<-1$ dex and halo-like kinematics, show a prominent, $\sim 0.15$ dex, depletion of [Mg/Fe] compared to field stars of the same metallicity. This may indicate their {\it ex situ} formation history.

\item NGC~2808 and NGC~1851 exhibit remarkably similar chemical abundance patterns and overlap in metallicity that reinforces the evidence for their common origin proposed in the literature.

\item Large intra-cluster spreads in [Mg/Fe], compared to the field population, are seen in the clusters M~2, NGC 2808, NGC 4833 and M15, corroborating with the long-postulated scenario that globular clusters have undergone multiple episodes of star formation and self-enrichment. On the other hand, the clusters are homogeneous in [Ti/Fe].

\item The metal-poor globular cluster NGC 4372 stands out in comparison with the other globular clusters with a similar metallicity. Its [Mg/Fe] spread is relatively small, consistent with the study by \citet{sanroman2015}. Given our standard abundance uncertainties of $\sim 0.1$ dex, which exceed the intra-cluster dispersion, the cluster is homogeneous in [Fe/H], [Mg/Fe] and [Ti/Fe].

\item For M15 and NGC 4833, which are the most metal-poor clusters in our sample, we find strong evidence for a multi-modality in [Mg/Fe]. However, our samples are too small to draw statistically robust conclusions on whether these clusters host two or more sub-populations. 

\end{itemize}

The combination of NLTE models and {\it the Payne} is a powerful tool for homogeneous analysis of the stellar parameters and chemical abundances. Our results for a large sample of stars in wide range of metallicity suggests that NLTE effects are significant for metal-poor regime ($\feh<-1$) and should be always taken into account.  
\begin{acknowledgements}
We thank Nikolay Kacharov, Diane Feuillet and David Hogg for valuable discussions. We thank anonymous referee for useful suggestions.
Based on data products from observations made with ESO Telescopes at the La Silla Paranal Observatory under programme ID 188.B-3002. This work has made use of data from the European Space Agency (ESA) mission {\it Gaia} (\url{https://www.cosmos.esa.int/gaia}), processed by the {\it Gaia} Data Processing and Analysis Consortium (DPAC, \url{https://www.cosmos.esa.int/web/gaia/dpac/consortium}). Funding for the DPAC has been provided by national institutions, in particular the institutions participating in the {\it Gaia} Multilateral Agreement. This research has made use of the WEBDA database, operated at the Department of Theoretical Physics and Astrophysics of the Masaryk University. We acknowledge support by the Collaborative Research centre SFB 881 (Heidelberg University) of the Deutsche Forschungsgemeinschaft (DFG, German Research Foundation). YST is supported by the NASA Hubble Fellowship grant HST-HF2-51425.001 awarded by the Space Telescope Science Institute.
\end{acknowledgements}



\begin{appendix}
\section{Supplementary tables}
In Table~\ref{tab:sist} we list the sensitivity of measured abundance ratios to the typical errors in atmospheric parameters. We run the analysis with one parameter fixed to a perturbed value, allowing code to fit the others. We list the quadratic sum of individual sensitivities in the last row as the total systematic error. The results are given for one star from the open cluster NGC 2243, two stars in the metal-rich cluster NGC 104 and two stars in metal-poor cluster M 15. 

In Table~\ref{tab:gbs}, we provide NLTE stellar parameters for the highest $\snr$ spectra of 19 Gaia benchmark stars along with reference values from \citet{Jofre2015}. The last row indicates the mean difference of our values. 

In Table~\ref{tab:gcinfo}, we list cluster parameters from the literature including equatorial coordinates, heliocentric distances, reddening, mean RV, age and $\feh$.

 In Table~\ref{tab:newnlte1}, we provide maximum likelihood estimates of the cluster average abundances and internal dispersions (more precisely, abundance dispersion of the cluster that is not accounted for by the abundance measurement systematic error) with associated errors.

In Table~\ref{tab:catalog}, we provide NLTE/LTE stellar parameters and abundances with systematic errors for all $742$ stars in the cluster sample.  

\begin{table}
	\centering
	\caption{Sensitivity of abundance ratios to errors in atmospheric parameters.}
	\label{tab:sist}
	\begin{tabular}{lccc} 
		\hline
		star/parameter&$\Delta$[Fe/H] &$\Delta$[Mg/Fe] &$\Delta$[Ti/Fe]\\
		& dex & dex & dex\\
		\hline
		\multicolumn{4}{c}{06291929-3125331~$\teff$=6689,~$\logg$=4.22,~$\feh$=-0.52}\\
		\hline
		$\teff$ +150~K& 0.08 & -0.01 &0.01\\
		$\logg$ +0.3~dex&0.07 & -0.06 &0.08\\
		$\feh$ +0.1~dex& $\cdots$  & -0.02 &0.01 \\
		$\Vmic$ +0.2~$\kms$& 0.01 & 0.02 &-0.01\\
		total & 0.10&0.06&0.07\\
		\hline
		\multicolumn{4}{c}{00225472-7203461~$\teff$=5146,~$\logg$=3.08,~$\feh$=-0.75}\\
		\hline
		$\teff$ +150~K& 0.11 & -0.07 &0.04\\
		$\logg$ +0.3~dex& 0.08 & -0.10 &0.08\\
		$\feh$ +0.1~dex& $\cdots$  & -0.05 &0.02\\
		$\Vmic$ +0.2~$\kms$&-0.03 & 0.02 &-0.04\\
		total & 0.14&0.14&0.10\\
		\hline
    	\multicolumn{4}{c}{00250332-7201108~$\teff$=4662,~$\logg$=2.21,~$\feh$=-0.78}\\
		\hline
		$\teff$ +150~K& 0.13 & -0.08 &0.03\\
		$\logg$ +0.3~dex& 0.08 & -0.09 &0.04\\
		$\feh$ +0.1~dex& $\cdots$ & -0.04 &-0.01\\
		$\Vmic$ +0.2~$\kms$&-0.04 & 0.02 &-0.02\\
		total & 0.16&0.12&0.05\\
		\hline
        \multicolumn{4}{c}{21300747+1210115~$\teff$=5150,~$\logg$=1.99,~$\feh$=-2.32}\\
		\hline
		$\teff$ +150~K& 0.10 & -0.05 &-0.01\\
		$\logg$ +0.3~dex& 0.01 & -0.01 &0.10\\
		$\feh$ +0.1~dex& $\cdots$  & -0.05 &-0.03\\
		$\Vmic$ +0.2~$\kms$&-0.04 &0.02 &-0.01\\
		total & 0.11&0.08&0.11\\
		\hline
		\multicolumn{4}{c}{21295615+1210296~$\teff$=5329,~$\logg$=2.30,~$\feh$=-2.26}\\
		\hline
		$\teff$ +150~K& 0.10 & -0.02 &0.02\\
		$\logg$ +0.3~dex& 0.06 & -0.01 &0.06\\
		$\feh$ +0.1~dex& $\cdots$  & -0.03 &0.01\\
		$\Vmic$ +0.2~$\kms$&0.06 &-0.01 & 0.03\\
		total & 0.13&0.03&0.07\\
		\hline
	\end{tabular}
\end{table}

\begin{table*}
	\centering
	\caption{Gaia benchmark stars parameters from NLTE fit (max $\snr$) and reference study \protect\citep{Jofre2015}, except where noted.}
	\label{tab:gbs}
	\begin{tabular}{lcccc} 
		\hline
Star & $\teff$,K & $\logg$,dex&$\feh$,dex&$\Vmic$, $\kms$\\
     &  fit, ref& fit, ref& fit, ref& fit, ref\\
		\hline
HD107328&4384, 4496 $\pm$  59&1.90, 2.09 $\pm$ 0.14&-0.60, -0.38 $\pm$ 0.16&1.71, 1.65 $\pm$ 0.26\\
HD220009&4336, 4275 $\pm$  54&1.86, 1.47 $\pm$ 0.14&-0.79, -0.79 $\pm$ 0.13&1.42, 1.49 $\pm$ 0.14\\
ksiHya&5045, 5044 $\pm$  38&3.01, 2.87 $\pm$ 0.02&-0.05, 0.11 $\pm$ 0.20&1.54, 1.40 $\pm$ 0.32\\
muLeo&4462, 4474 $\pm$  60&2.45, 2.51 $\pm$ 0.09&0.01, 0.20 $\pm$ 0.15&1.54, 1.28 $\pm$ 0.26\\
HD122563&4771, 4636 $\pm$  37\tablefootmark{1}&1.29, 1.42 $\pm$ 0.01\tablefootmark{2}&-2.56, -2.52 $\pm$ 0.11\tablefootmark{3}&2.53, 1.92 $\pm$ 0.11\\
HD140283&5888, 5787 $\pm$  48\tablefootmark{1}&3.63, 3.57 $\pm$ 0.12&-2.39, -2.34 $\pm$ 0.03\tablefootmark{3}&2.16, 1.56 $\pm$ 0.20\\
delEri&5006, 4954 $\pm$  26&3.61, 3.75 $\pm$ 0.02&-0.00, 0.01 $\pm$ 0.05&1.15, 1.10 $\pm$ 0.22\\
epsFor&5070, 5123 $\pm$  78&3.28, 3.52 $\pm$ 0.07&-0.65, -0.65 $\pm$ 0.10&1.14, 1.04 $\pm$ 0.13\\
18Sco&5838, 5810 $\pm$  80&4.32, 4.44 $\pm$ 0.03&0.02, -0.02 $\pm$ 0.03&1.27, 1.07 $\pm$ 0.20\\
alfCenB&5167, 5231 $\pm$  20&4.33, 4.53 $\pm$ 0.03&0.14, 0.17 $\pm$ 0.10&1.06, 0.99 $\pm$ 0.31\\
muAra&5743, 5902 $\pm$  66&4.05, 4.30 $\pm$ 0.03&0.22, 0.30 $\pm$ 0.13&1.32, 1.17 $\pm$ 0.13\\
betVir&6259, 6083 $\pm$  41&4.06, 4.10 $\pm$ 0.02&0.18, 0.19 $\pm$ 0.07&1.51, 1.33 $\pm$ 0.09\\
epsEri&5079, 5076 $\pm$  30&4.54, 4.60 $\pm$ 0.03&-0.14, -0.14 $\pm$ 0.06&1.11, 1.14 $\pm$ 0.05\\
etaBoo&6183, 6099 $\pm$  28&3.84, 3.80 $\pm$ 0.02&0.27, 0.27 $\pm$ 0.08&1.52, 1.52 $\pm$ 0.19\\
HD22879&5907, 5868 $\pm$  89&3.98, 4.27 $\pm$ 0.03&-0.80, -0.91 $\pm$ 0.05&1.24, 1.05 $\pm$ 0.19\\
HD49933&6718, 6635 $\pm$  91&4.16, 4.20 $\pm$ 0.03&-0.36, -0.46 $\pm$ 0.08&1.51, 1.46 $\pm$ 0.35\\
HD84937&6481, 6356 $\pm$  97&3.91, 4.15 $\pm$ 0.06&-2.00, -1.99 $\pm$ 0.02\tablefootmark{3}&1.76, 1.39 $\pm$ 0.24\\
Procyon&6686, 6554 $\pm$  84&3.91, 3.99 $\pm$ 0.02&0.03, -0.04 $\pm$ 0.08&1.83, 1.66 $\pm$ 0.11\\
tauCet&5349, 5414 $\pm$  21&4.26, 4.49 $\pm$ 0.01&-0.52, -0.54 $\pm$ 0.03&1.00, 0.89 $\pm$ 0.28\\
		\hline
		<ref-fit>&-29 $\pm$ 88& 0.09 $\pm$ 0.16& 0.02 $\pm$ 0.09&-0.16 $\pm$ 0.18\\
		\hline
	\end{tabular}
		\tablefoot{
 In order to be consistent with our reference solar [Fe/H] scale, we subtracted $0.05$~dex
 from \protect\citet{Jofre2015} and $0.03$~dex from \protect\citet{amarsi2016} metallicities.\\
 References:\tablefoottext{1}{\protect\citet{Karovicova2018}}
\tablefoottext{2}{\protect\citet{creevey2019}}
\tablefoottext{3}{\protect\citet{amarsi2016}}}

\end{table*}

\begin{table*}
	\centering
	\caption{Cluster parameters used in this work.	}
	\label{tab:gcinfo}
	\begin{tabular}{lccccccc} 
		\hline
Cluster & $\alpha$,deg & $\delta$,deg&$d_{\odot}$, kpc&E(B-V)&<RV>, $\kms$ &Age, Gyr&$\feh$,dex\\
		\hline
NGC 3532 (oc)& 166.4125 & -58.7533 & 0.5\tablefootmark{1}& 0.03\tablefootmark{1}& 4.3 & 0.3 & -0.01 \\
NGC 5927 (gc)& 232.0029 & -50.6730 & 7.7&0.45&-100.5 & 11.9 & -0.48 \\
NGC 2243 (oc)&  97.3917 & -31.2833 &4.5 &0.05&59.8 & 3.8\tablefootmark{2} & -0.57\tablefootmark{2} \\
NGC 104 (gc)&   6.0224 & -72.0815 &4.5\tablefootmark{3}&0.04& -18.7 & 12.5 & -0.75 \\
NGC 1851 (gc)&  78.5281 & -40.0465 & 12.1&0.02&320.9 & 10.5 & -1.10 \\
NGC 2808 (gc)& 138.0129 & -64.8635 & 9.6&0.22&102.8 & 10.9 & -1.14 \\
NGC 362 (gc)&  15.8094 & -70.8488 & 8.5\tablefootmark{3}&0.05&222.9 & 10.9 & -1.23 \\
M 2 (gc)& 323.3626 & -0.8233 & 11.5&0.06&-6.7 & 12.0 & -1.52 \\
NGC 6752 (gc)& 287.7170 & -59.9846 &4.0&0.02& -27.4 & 12.3 & -1.43 \\
NGC 1904 (gc)&  81.0441 & -24.5242 & 12.9&0.01&205.8 & 11.1 & -1.37 \\
NGC 4833 (gc)& 194.8913 & -70.8765 &6.6&0.32& 201.1 & 12.7 & -1.97 \\
NGC 4372 (gc)& 186.4393 & -72.6591 &5.8 &0.30..0.80\tablefootmark{4}&72.6 & 12.5 & -1.88 \\
M 15 (gc)& 322.4930 & 12.1670 & 10.4&0.10&-106.6 & 13.0 & -2.25 \\

		\hline
	\end{tabular}
	\tablefoot{
	 The coordinates and radial velocities from SIMBAD database, ages and $\feh$ from \protect\citet{kruijssen2018} for globular clusters (gc) and WEBDA database for open clusters (oc), distances and E(B-V) are from \protect\citet[][2010 edition]{harris} (gc) or WEBDA (oc) databases, except where noted. \\
	 References:
\tablefoottext{1}{\protect\citet{Fritzewski2019}}
\tablefoottext{2}{\protect\citet{Anthony-Twarog2005}}
\tablefoottext{3}{\protect\citet{Chen2018}}
\tablefoottext{4}{\protect\citet{Kacharov2014}}

}
\end{table*}

\begin{table*}[tp]\small
    \centering
    \caption{Maximum likelihood estimates of the cluster average abundances and internal dispersions.}
    \begin{tabular}{lcccccc}
    \hline
    Cluster  & <$\feh_{\rm NLTE}$> & $\sigma\feh_{\rm NLTE}$ &  <$\mgfe_{\rm NLTE}$> & $\sigma\mgfe_{\rm NLTE}$ & <$\tife_{\rm NLTE}$> & $\sigma\tife_{\rm NLTE}$ \\
     N stars & <$\feh_{\rm LTE}$> & $\sigma\feh_{\rm LTE}$ &  <$\mgfe_{\rm LTE}$> & $\sigma\mgfe_{\rm LTE}$ & <$\tife_{\rm LTE}$> & $\sigma\tife_{\rm LTE}$ \\
     & dex & dex &dex &dex &dex &dex \\
    \hline
NGC 3532 &-0.10 $\pm$ 0.02&0.00 $\pm$ 0.02&-0.09 $\pm$ 0.03&0.00 $\pm$ 0.03&0.01 $\pm$ 0.03&0.00 $\pm$ 0.03\\
12&-0.09 $\pm$ 0.02&0.00 $\pm$ 0.03&-0.07 $\pm$ 0.03&0.00 $\pm$ 0.03&0.01 $\pm$ 0.03&0.00 $\pm$ 0.03\\
\hline
NGC 5927 &-0.48 $\pm$ 0.02&0.00 $\pm$ 0.02&0.39 $\pm$ 0.02&0.00 $\pm$ 0.02&0.29 $\pm$ 0.01&0.00 $\pm$ 0.02\\
47&-0.49 $\pm$ 0.02&0.00 $\pm$ 0.02&0.41 $\pm$ 0.02&0.00 $\pm$ 0.02&0.23 $\pm$ 0.01&0.00 $\pm$ 0.02\\
\hline
NGC 2243 &-0.52 $\pm$ 0.01&0.00 $\pm$ 0.01&0.15 $\pm$ 0.01&0.00 $\pm$ 0.02&0.02 $\pm$ 0.01&0.00 $\pm$ 0.04\\
84&-0.57 $\pm$ 0.01&0.00 $\pm$ 0.02&0.26 $\pm$ 0.01&0.00 $\pm$ 0.02&0.01 $\pm$ 0.01&0.00 $\pm$ 0.05\\
\hline
NGC 104 &-0.74 $\pm$ 0.02&0.00 $\pm$ 0.02&0.38 $\pm$ 0.02&0.00 $\pm$ 0.02&0.30 $\pm$ 0.01&0.00 $\pm$ 0.02\\
68&-0.75 $\pm$ 0.02&0.00 $\pm$ 0.02&0.42 $\pm$ 0.02&0.00 $\pm$ 0.02&0.26 $\pm$ 0.01&0.00 $\pm$ 0.03\\
\hline
NGC 1851 &-1.11 $\pm$ 0.01&0.00 $\pm$ 0.02&0.22 $\pm$ 0.01&0.00 $\pm$ 0.02&0.28 $\pm$ 0.01&0.00 $\pm$ 0.02\\
88&-1.15 $\pm$ 0.01&0.00 $\pm$ 0.02&0.36 $\pm$ 0.01&0.00 $\pm$ 0.02&0.24 $\pm$ 0.01&0.00 $\pm$ 0.01\\
\hline
NGC 2808 &-1.01 $\pm$ 0.03&0.00 $\pm$ 0.03&0.11 $\pm$ 0.03&0.09 $\pm$ 0.03&0.33 $\pm$ 0.02&0.00 $\pm$ 0.02\\
25&-1.03 $\pm$ 0.03&0.00 $\pm$ 0.03&0.22 $\pm$ 0.03&0.00 $\pm$ 0.09&0.30 $\pm$ 0.01&0.00 $\pm$ 0.02\\
\hline
NGC 362 &-1.05 $\pm$ 0.02&0.00 $\pm$ 0.02&0.15 $\pm$ 0.01&0.00 $\pm$ 0.02&0.29 $\pm$ 0.01&0.00 $\pm$ 0.01\\
62&-1.09 $\pm$ 0.02&0.00 $\pm$ 0.02&0.26 $\pm$ 0.02&0.00 $\pm$ 0.02&0.26 $\pm$ 0.01&0.00 $\pm$ 0.01\\
\hline
M 2 &-1.47 $\pm$ 0.01&0.00 $\pm$ 0.02&0.17 $\pm$ 0.01&0.07 $\pm$ 0.01&0.23 $\pm$ 0.01&0.00 $\pm$ 0.02\\
78&-1.54 $\pm$ 0.01&0.00 $\pm$ 0.02&0.34 $\pm$ 0.02&0.08 $\pm$ 0.02&0.25 $\pm$ 0.01&0.00 $\pm$ 0.01\\
\hline
NGC 6752 &-1.48 $\pm$ 0.01&0.00 $\pm$ 0.01&0.20 $\pm$ 0.01&0.03 $\pm$ 0.02&0.17 $\pm$ 0.01&0.00 $\pm$ 0.01\\
110&-1.56 $\pm$ 0.01&0.00 $\pm$ 0.01&0.35 $\pm$ 0.01&0.04 $\pm$ 0.02&0.23 $\pm$ 0.01&0.00 $\pm$ 0.01\\
\hline
NGC 1904 &-1.51 $\pm$ 0.02&0.00 $\pm$ 0.02&0.16 $\pm$ 0.01&0.04 $\pm$ 0.02&0.21 $\pm$ 0.01&0.00 $\pm$ 0.02\\
44&-1.60 $\pm$ 0.02&0.00 $\pm$ 0.02&0.31 $\pm$ 0.02&0.00 $\pm$ 0.04&0.24 $\pm$ 0.01&0.00 $\pm$ 0.03\\
\hline
NGC 4833 &-1.88 $\pm$ 0.02&0.00 $\pm$ 0.02&0.18 $\pm$ 0.03&0.15 $\pm$ 0.02&0.22 $\pm$ 0.01&0.00 $\pm$ 0.02\\
33&-2.08 $\pm$ 0.02&0.00 $\pm$ 0.03&0.36 $\pm$ 0.03&0.18 $\pm$ 0.03&0.24 $\pm$ 0.02&0.00 $\pm$ 0.02\\
\hline
NGC 4372 &-2.07 $\pm$ 0.02&0.00 $\pm$ 0.02&0.31 $\pm$ 0.01&0.00 $\pm$ 0.03&0.20 $\pm$ 0.01&0.00 $\pm$ 0.02\\
45&-2.33 $\pm$ 0.02&0.00 $\pm$ 0.02&0.51 $\pm$ 0.02&0.00 $\pm$ 0.04&0.22 $\pm$ 0.01&0.00 $\pm$ 0.02\\
\hline
M 15 &-2.28 $\pm$ 0.02&0.00 $\pm$ 0.02&0.22 $\pm$ 0.03&0.16 $\pm$ 0.02&0.21 $\pm$ 0.02&0.00 $\pm$ 0.02\\
46&-2.58 $\pm$ 0.02&0.00 $\pm$ 0.02&0.36 $\pm$ 0.04&0.22 $\pm$ 0.03&0.19 $\pm$ 0.01&0.00 $\pm$ 0.02\\
\hline
    \end{tabular}

    \label{tab:newnlte1}
\end{table*}

\begin{table*}[tp]\small
    \centering
    \caption{NLTE and LTE stellar parameters, abundance ratios with uncertainties for the 742 stars in clusters sample.}
    \begin{tabular}{ccccccccccccc}
    \hline
    Star &Cluster &$\snr$&\multicolumn{2}{c}{$\teff$}&\multicolumn{2}{c}{$\logg$}&\multicolumn{2}{c}{$\Vmic$}&\multicolumn{2}{c}{$\feh$}&\multicolumn{2}{c}{$\Delta\feh$}\\
        & & & NLTE & LTE & NLTE &  LTE& NLTE &  LTE& NLTE &  LTE& NLTE &  LTE\\
       HHMMSSss-DDMMSSs& & \AA$^{-1}$& K & K & dex & dex &$\kms$&$\kms$& dex&dex& dex&dex\\
       \hline
       11070908-5839114 &NGC 3532 &537.5 & 5389 & 5379& 4.53 &4.52&1.39&1.28&-0.10&-0.09&0.09&0.08\\
       11071871-5849027 &NGC 3532 &521.9 & 5606 & 5615& 4.54& 4.54& 1.43& 1.41&-0.08&-0.08&0.09&0.09 \\
     ..& .. & .. & .. &.. & .. &.. & ..& ..& ..& ..& ..& ..\\  
     \hline
      & & & & & & &  & & &  & & \\  
     \cline{2-13}
    &\multicolumn{2}{c}{$\mgfe$}&\multicolumn{2}{c}{$\Delta\mgfe$}&\multicolumn{2}{c}{$\tife$}&\multicolumn{2}{c}{$\Delta\tife$}&\multicolumn{2}{c}{$\mnfe$}&\multicolumn{2}{c}{$\Delta\mnfe$}\\
    & NLTE & LTE & NLTE &  LTE& NLTE &  LTE& NLTE &  LTE& NLTE &  LTE& NLTE & LTE \\
    & dex&dex& dex&dex& dex&dex& dex&dex& dex&dex& dex&dex\\
    \cline{2-13}
    &-0.08 &-0.08 & 0.09 &0.10 &-0.01 &0.04 & 0.08 &0.14 &-0.14 &-0.26 & 0.11 &0.13\\
    &-0.08 &-0.07 & 0.09 &0.10 & 0.02 &0.03 & 0.06 &0.10 &-0.16 &-0.26 & 0.12 &0.12\\
    & .. & .. & .. &.. & .. &.. & ..& ..& ..& ..& ..& ..\\
    \cline{2-13}

         \end{tabular}
         \tablefoot{The full table is available at the CDS.}

    \label{tab:catalog}
\end{table*}

\section{Instrumental profile}
\label{LSF}
\begin{figure}
\includegraphics[width=\columnwidth]{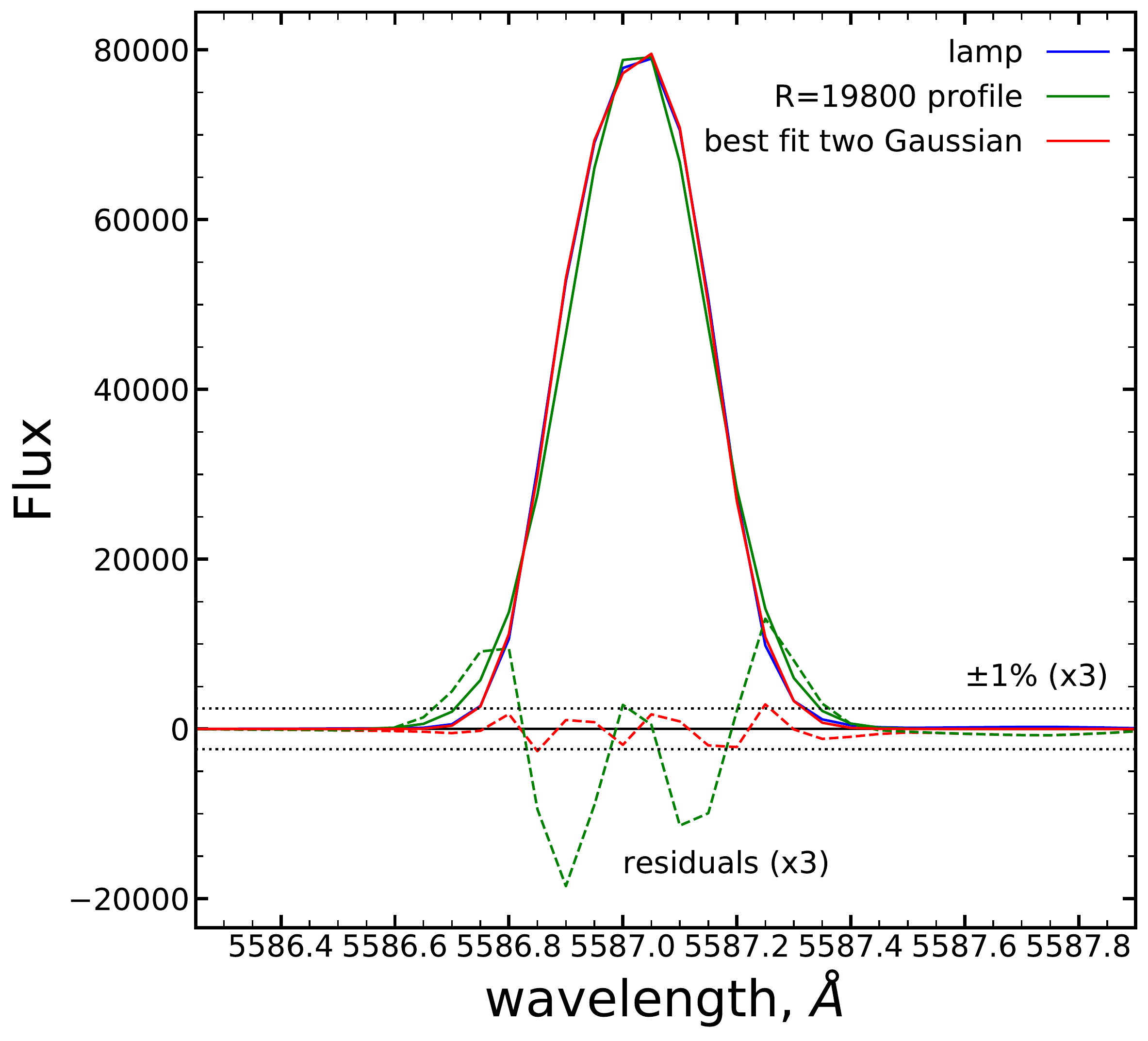}
\caption{Results of instrumental profile fitting. Residuals of the fit are up-scaled three times.  }
\label{fig:lsfplot} 
\end{figure}

Similarly to the technique that \citet{Damiani2016} used to obtain an instrumental profile for the Giraffe HR15N setting, we used a sum of two Gaussian profiles to fit the line at 5578~\AA ~in the calibration spectrum of the thorium-argon lamp, downloaded from the ESO webpage\footnote{\url{http://www.eso.org/observing/dfo/quality/GIRAFFE/pipeline/SKY/html/GI_SRBS_2004-09-26T22_48_10.511_Medusa2_H548.8nm_o10.fits}}. It is shown in Fig. \ref{fig:lsfplot} that such a new instrumental profile describes the spectral profile much better than a single Gaussian computed according to the reported resolution of HR10 setting $R=19\,800$. The error of a single Gaussian profile can be up to 5-7\% while, using two-Gaussian the profile error is alway below the 1\% level. The resulting instrumental profile with the best-fitted parameters is listed below:
\begin{equation}
    \lambda(v)=\frac{A_1}{\sqrt{2 \pi \sigma_{1}^2}} {\rm exp} \left (-\frac{(v-v_1)^2}{2 \sigma_{1}^2} \right)+\frac{A_2}{\sqrt{2 \pi \sigma_{2}^2}} {\rm exp} \left (-\frac{(v-v_2)^2}{2 \sigma_{2}^2} \right)
\end{equation}
with $A_1=0.465$, $A_2=0.194$, $\sigma_1= 4.971\kms$, $\sigma_2 =3.799\kms$, $v_1=-2.249\kms$, $v_2= 5.754\kms$.

\section{Kinematic assignment of the populations.}
\label{distances}

We employ the cluster distances listed in Table~\ref{tab:gcinfo}. They were obtained from the colour magnitude diagram horizontal branch (globular clusters \citet[][2010 edition]{harris}) or turn-off point (open clusters WEBDA database) fitting. The same distance is assumed for all stars within given cluster. We also take proper motions from Gaia DR2 \citep{gdr2} and radial velocities from our analysis and compute galactocentric rectangular velocity components (U,V,W) for all stars in the clusters, using \emph{Astropy} \citep{astropy}, package, with respect to solar motion from \citet{schoenrich2010}.

The computed velocities are used to calculate the probability ratios $TD/D$ and  $TD/H$ \citep[][ Appendix 1]{Bensby2014}, which allow us to assign population membership to the clusters. We use the following selection criteria: thick disk if  $TD/D > 2$ and $TD/H>~1$; thin disk if $TD/D< 0.5$; halo if $TD/H<1$.  Only the open cluster NGC~2243 has a probability ratio of $TD/D=1.25$ in between the thin and the thick disk. We therefore decide to assign it to the thick disk on the basis of its large separation ($|z|=1$~kpc) from the Galactic plane.  

\end{appendix}

\end{document}